\documentclass[5p,times]{elsarticle}

\usepackage[utf8]{inputenc}
\usepackage{mathtools}
\usepackage{bm}

\usepackage{enumitem}
\usepackage[colorlinks=true, allcolors=blue]{hyperref}

\usepackage{amsmath,amssymb,amsfonts}
\usepackage{algorithmic}
\usepackage{graphicx}
\usepackage{textcomp}
\usepackage[dvipsnames]{xcolor}
\usepackage{listings}
\usepackage[ruled,vlined,linesnumbered,noend]{algorithm2e}

\usepackage{subfigure}
\usepackage{subcaption}
\usepackage{combelow}
\usepackage{multirow}

\emergencystretch=\maxdimen
\newlength{\commentWidth}
\newcommand{\atcp}[1]{\tcp*[r]{\makebox[\commentWidth]{#1\hfill}}}

\begin{document}

\begin{frontmatter}

\title{\textsc{CleanNews}: a Network-aware Fake News Mitigation Architecture for Social Media}

\author[1]{Maria-Diana Cotelin\fnref{c2}}
\ead{maria\_diana.cotelin@stud.acs.upb.ro}

\author[1,2]{Ciprian-Octavian~Truică\corref{c1}\fnref{c2}}
\ead{ciprian.truica@upb.ro}

\author[1,2]{Elena-Simona Apostol\fnref{c2}}
\ead{elena.apostol@upb.ro}

\cortext[c1]{Corresponding author.}
\fntext[c2]{These authors contributed equally to this article.}

\affiliation[1]{organization={Computer Science and Engineering Department, Faculty of Automatic Control and Computers, National University of Science and Technology Politehnica Bucharest},
    addressline={Splaiul Independentei 313}, 
    city={Bucharest},
    postcode={060042}, 
    country={Romania}
}

\affiliation[2]{organization={Academy of Romanian Scientists},
    addressline={Ilfov 3}, 
    city={Bucharest},
    postcode={050044},     
    country={Romania}
}

\begin{abstract}
With the widespread use of the internet and handheld devices, social media now holds a power similar to that of old newspapers.
People use social media platforms for quick and accessible information. However, this convenience comes with a variety of risks.
Anyone can freely post content, true or false, with the probability of remaining online forever.
This makes it crucial to identify and tackle misinformation and disinformation on online platforms.
In this article, we propose \textsc{CleanNews}, a comprehensive architecture to identify fake news in real-time accurately.
\textsc{CleanNews} uses advanced deep learning architectures, combining convolutional and bidirectional recurrent neural networks, i.e., LSTM and GRU, layers to detect fake news.
A key contribution of our work is a novel embedding technique that fuses textual information with user network structure, allowing the model to jointly learn linguistic and relational cues associated with misinformation.
Furthermore, we use two network immunization algorithms, i.e., SparseShield and NetShield, to mitigate the spread of false information within networks.
We conduct extensive ablation studies to evaluate the contribution of each model component and systematically tune hyperparameters to maximize performance.
The experimental evaluation on two real-world datasets shows the efficacy of \textsc{CleanNews} in combating the spread of fake news.
\end{abstract}


\end{frontmatter}

\section{Introduction}

Many people get their information from social networks, often involuntarily due to the time spent on apps and the ease of access.
However, this comfort comes with significant risks.
Anyone can publish content, regardless of its authenticity, and the content can remain online for an extended period.
This unregulated flow of information has led to serious consequences.
Russia's use of bans on ``fake news'' during the war against Ukraine created a mythologized reality about ``special military operation'' to enable the government to control public debate and misrepresent the war.
This conditional truth significantly threatens freedom of expression and democracy in Russia, helping the government maintain public support for the war~\cite{sherstoboeva2024russian}. 
The Israeli-Palestine conflict, especially during the 2023 Israel-Hamas war, has been another major example of how fake news and disinformation shape public perception, policy reactions, and even violence offline.
Shortly after Hamas's 7 October 2023 attack, Israeli officials and media reported that 40 babies were beheaded.
The claim made headlines worldwide and fueled outrage, but was later retracted or left unverified, leading to accusations of disinformation~\cite{Reis2024}.

This article presents a complete solution designed to efficiently detect and mitigate the spread of fake news in real-time.
Our proposed system has three key modules: the Preprocessing Module, the Detection Module, and the Mitigation Module.
To guide our work, we address the following research questions:
\begin{itemize}
    \item \textbf{RQ1:} How can textual data from social media be preprocessed to enhance the performance of fake news detection models?
    \item \textbf{RQ2:} What are the key components and design steps required to construct an effective deep learning pipeline (including feature extraction, model architecture, and optimization) for accurately detecting fake news in dynamic online environments?
    \item \textbf{RQ3:} Which network-based strategies can be applied to effectively mitigate the spread of misinformation once it has been identified?
\end{itemize}

The Preprocessing Module uses well-known techniques to clean and refine the datasets by removing unnecessary information. 
This process involves several steps, such as: 
1) \textbf{Data Cleaning}, which involves removing irrelevant data (e.g., special characters, extra whitespace, etc.),
2) \textbf{Stopword Removal}, which involves removing words that do not carry significant meaning, such as ``and'', ``the'', are removed from the text,
3) \textbf{Tokenization}, which involves splitting the text into individual words or tokens, and
4) \textbf{Stemming and Lemmatization}, which employs methods that extract the root forms of words.

The Detection Module uses advanced deep learning neural architectures to accurately identify fake news on social media platforms.
It analyzes the content and patterns of information to accurately identify false or misleading content.
In this module, we used a combination of Convolutional Neural Networks (CNNs) and variants of Recurrent Neural Networks (RNNs), specifically Bidirectional Long Short-Term Memory (BiLSTM), and Bidirectional Gated Recurrent Unit (BiGRU) layers to analyze and process the data.
The CNN layers are used for capturing spatial features, while the BiLSTM and BiGRU layers are used for understanding temporal dependencies and sequential patterns in the data.

The Mitigation Module then applies targeted strategies to limit the spread of identified false news within social networks.
Using results from the Preprocessing and Detection Modules, it targets nodes in the network that spread false information. 
Network Immunization strategies, i.e., SparseShield~\cite{Petrescu2021} and NetShield~\cite{Chen2015NetShield}, are deployed to reduce the propagation of fake news within a network.

This work makes the following key contributions:
\begin{itemize}
    \item We propose a three-stage modular architecture for real-time fake news detection and mitigation, integrating preprocessing, deep learning-based detection, and network-based intervention strategies.
    \item We develop and evaluate a hybrid deep learning model combining CNN, BiLSTM, and BiGRU layers, which improves the accuracy and robustness of fake news classification on social media datasets.
    \item We apply graph-based mitigation techniques (SparseShield and NetShield) to minimize the spread of false information by targeting the most influential nodes in a network.
\end{itemize}

This work is structured as follows:
Section~\ref{sec:sota} provides insights into the recent literature relevant to the topic.
Section~\ref{sec:methodology} details the architecture of \textsc{CleanNews}.
Section~\ref{sec:results} offers an experimental evaluation of our proposed solution.
Section~\ref{sec:discussion} discusses our findings and their limitations.
Finally, Section~\ref{sec:conclusions} presents the conclusions of this work and outlines potential directions for future research.

\section{Related Work}\label{sec:sota}

In recent years, research on the detection of misleading content has advanced significantly.
This is due to improvements in natural language processing and graph-based analysis of online networks.

In the area of text representation, the current literature has explored a variety of embedding techniques that capture semantic meaning.
Traditional word embeddings such as Word2Vec, FastText, and GloVe are commonly used ~\cite{Ilie2021, mallik2024word2vec, taher2022automatic,abualigah2024fake}, while more recent approaches focus on transformer-based embeddings~\cite{Cotelin2025Exist,Petrescu2024,Truica2022Misrobaerta} and mixtures of transformes~\cite{Petrescu2024,Petrescu2025bestofclef}  for richer contextual representation.
Sentence-level representations using transformer models have also shown promise in detecting subtle and context-dependent misinformation~\cite{Truica2022checktaht2022}.
At a higher level of granularity, document embeddings have been employed to represent entire articles or social media posts for classification tasks~\cite{Truica2023}.

When it comes to models, various deep learning architectures are employed as classifiers for the detection of harmful content.
State-of-the-art solutions frequently rely on Recurrent Neural Networks (RNNs), particularly LSTMs (Long Short-Term Memory), GRUs (Gated Recurrent Units), due to their ability to model sequential dependencies in text~\cite{Badjatiya2017deep}.
CNNs (Convolutional Neural Networks) are also widely used to capture local textual characteristics, especially short-form content such as tweets or comments~\cite{Badjatiya2017deep, Ilie2021, Truica2022Misrobaerta}.
Other approaches leverage both news content and social contexts, using Transformer architectures, specifically a Bidirectional and Autoregressive Transformer (BART) architecture~\cite{raza2022fake,tran2022ur,Truica2022Misrobaerta}, or variations of Bidirectional Encoder Representations from Transformers (BERT), such as DistilBERT, RoBERTa, and ALBERT~\cite{farhangian2024fake,szczepanski2021new,kaliyar2021fakebert}.
Other studies investigate the potential of Large Language Models (LLMs) in fake news detection, i.e., GPT 2~\cite{dhiman2024gbert},  GPT 3.5~\cite{hu2024bad,anirudh2023multilingual,anjos2023investigating}, or GPT 4o~\cite{roumeliotis2025fake}.
Other articles used LLM-empowered news reframings to inject style diversity into the training process~\cite{wu2024fake, askarizade2025enhancing}.

Another promising technique leverages Graph Convolutional Networks (GCNs), utilizing the inherent structure of social networks to enhance information modeling.
Specifically, this method adopts a tree-structured GCN to facilitate the hierarchical propagation of information, starting from the root tweet and extending through its associated comments.
To further strengthen the representation learning, a co-attention mechanism is integrated, allowing for the dynamic fusion of two key feature types: the textual semantics of the content and the underlying propagation structure of the posts.
This combined approach not only boosts the overall performance of the model but also contributes to improved training efficiency~\cite{zhang2024gbca,liu2024rumor,meel2021fake,wang2020fake,Lu2020GCAN,qian2021knowledge,ahammad2024roberta}.

Beyond detection, network immunization has appeared as a strategy to limit the spread of misinformation within social networks. 
These solutions focus on analyzing how harmful content propagates online and identifying intervention points.
The current literature has proposed multiple solutions that analyze the spread of content online~\cite{Petrescu2019, petrescu2023edsaensemble,Truica2021c}.
Some strategies focus on strengthening the network's resilience to misinformation by proactively identifying and targeting influential nodes in advance~\cite{Petrescu2021,Apostol2024contain}.
Others employ contra-active strategies that react to misinformation after it begins to spread~\cite{Truica2023MCWDST}.
A more recent and comprehensive solution, \textsc{StopHS}~\cite{Truica2024}, builds on top of the current literature and proposes a novel deep-learning architecture that combines textual analysis with network topology features to not only detect harmful content but also suppress its amplification across social platforms.
Moreover, ContCommRTD~\cite{Apostol2024ContCommRTD}, a distributed content-based community detection system for real-time disaster reporting, relies on detecting fake news to minimize its harmful effects during calamities.

\begin{figure*}[!htbp]
  \centering
  \includegraphics[width=1.60\columnwidth]{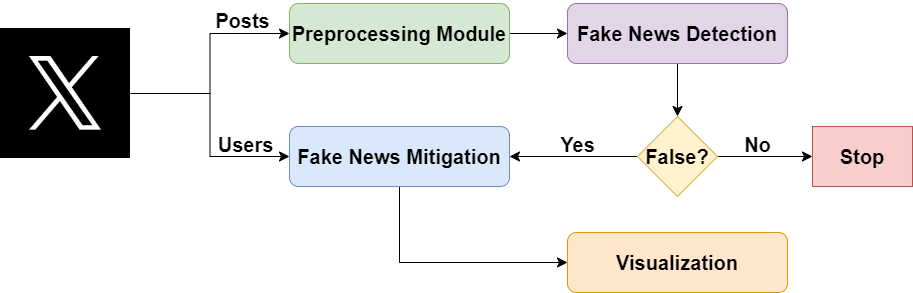}
  \caption{\textsc{CleanNews} architecture}
  \label{fig:pielinesolution}
\end{figure*}

One known limitation of this topic is related to the labeled datasets.
There are a limited number of such datasets, and they are often in English and focused on political news.
This affects the development and evaluation of detection models.
The lack of large-scale, diverse, and real-world datasets is a significant challenge~\cite{huoverview}.
While the use of LLMs to generate contextually relevant synthetic rumor instances addresses the class imbalance problem, very common in rumor detection, it has a major limitation in scalability for extremely large datasets.
This is due to the significant memory and processing power required.
The quality of generated synthetic data for augmentation may vary depending on the dataset.
This variation could affect performance in certain cases~\cite{askarizade2025enhancing}.
Another problem is that most studies are limited to content-based features, excluding other types of features like user-based or social context-based features.
The performance evaluation is primarily based on accuracy, which is acknowledged as a limitation~\cite{capuano2023content}.

\section{Methodology}\label{sec:methodology}

Figure~\ref{fig:pielinesolution} presents the \textsc{CleanNews}'s architecture.
Using a three-stage process (Preprocessing, Detection, and Mitigation), the proposed architecture is designed to identify the authenticity of a given post.
If the post is false, it proceeds to the immunization step.
Several advanced algorithms and models are employed within this framework:

\begin{itemize}
    \item \textbf{Preprocessing}: Use advanced models like DeBERTa~\cite{he2021deberta} and Node2Vec~\cite{Grover2016}.
    \item \textbf{Fake News Detection}: Apply BiGRU, BiLSTM, and CNN layers for identifying false posts.
    \item \textbf{Fake News Mitigation (Immunization)}: Includes strategies such as NetShield, SparseShield, and Random Solver for mitigating the spread of false information.
\end{itemize}

\subsection{Preprocessing Module}
The preprocessing module cleans the text and minimizes the vocabulary~\cite{Truica2016a} while preserving meaning~\cite{Truica2016Cats}.
It has four main steps:
\begin{enumerate}
    \item \textbf{Punctuation Marks, URL, Emoji, and Stop Words Removal}: 
    This step involves removing any punctuation marks, URLs, emojis, and common stop words from the text.
    Punctuation marks and emojis are not relevant for text detection.
    URLs are also removed, as they typically do not provide useful information for text analysis.
    Stop words are common words, e.g., ``of", ``the", etc., that are also removed because they do not provide significant semantic content.

    \item \textbf{Lemmatization}: 
    This step breaks the words down to their root meaning (lemma) to identify similarities. 
    For example, the words ``dancing", ``danced", ``dancer", and ``dances" all have the lemma ``dance``.
    This process helps to normalize the text and reduce the size of the vocabulary by treating different forms of a word as a single entity.

    \item \textbf{Padding/Truncating}: 
    In this step, the text is either padded with additional tokens or truncated to ensure that all input sequences have the same length. This is important for batch processing in neural networks because it guarantees that each input sequence has a uniform length.
    Padding typically involves adding special tokens (in our case $[PAD]$, from DeBERTa configuration) to shorter sequences, while truncating involves cutting off longer sequences to the expected length.

    \item \textbf{Word Encoding Using Word Embeddings}:
    The cleaned and normalized text is converted into numerical representations using word embeddings.
    These are dense vector representations of words that capture semantic meanings and relationships between words.
    By using pre-trained embeddings such as GloVe or embeddings from models like DeBERTa, each word in the text is mapped to a high-dimensional vector that captures its context and meaning.
\end{enumerate}

\begin{figure*}[!htbp]
  \centering
  \includegraphics[width=2\columnwidth]{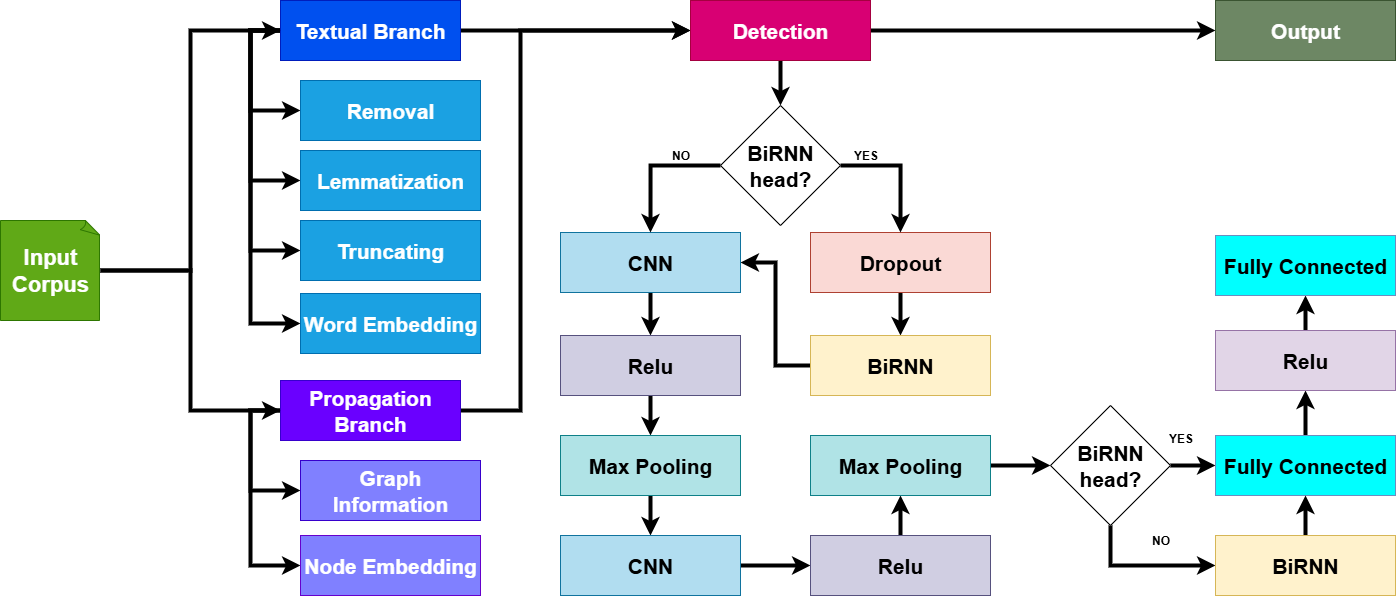}
  \caption{\textsc{CleanNews} Detection Architecture}
  \label{fig:CleanNewsDetArch}
\end{figure*}

\subsection{Fake News Detection}

The Detection Module uses four models as classifiers for each post.
These models are capable of analyzing the textual content of posts and making predictions regarding their truth.
The posts are classified as true, false, non-rumor, or unverified based on learned patterns and features. 
By combining Convolutional Neural Networks (CNN) and Bidirectional Long Short-Term Memory networks (BiLSTM) / Bidirectional Gated Recurrent Unit (BiGRU), the models capture both local and global dependencies in the text. 
This allows them to better understand subtle characteristics of different types of posts (see Table~\ref{table:models} and Figure~\ref{fig:CleanNewsDetArch}).

\begin{table}[!htbp]
\centering
\caption{\textsc{CleanNews} Detection Models}
\label{table:models}
\begin{tabular}{ll} 
 \hline
 \textbf{Model} & \textbf{Layers} \\ 
 \hline
BiRNN\_CNN & BiLSTM\_CNN\\ 
 & BiGRU\_CNN\\
CNN\_BiRNN & CNN\_BiLSTM\\
 & CNN\_BiGRU\\
 \hline
\end{tabular}
\end{table}

An \textit{Embedding} layer is used at the start of the models to store word embeddings concatenated with node embeddings. 
The word embeddings are obtained by using a pre-trained model. 
The node embeddings are trained on the propagation graphs.

We used a \textit{BiRNN} architecture because the data are in the form of word sequences. 
This architecture, which includes \textit{BiLSTM} and \textit{BiGRU} layers, is well-suited for this task.
The \textit{BiLSTM} layer is good at capturing long-term dependencies in text.
Another reason for using this layer is that it combines two \textit{LSTM} models that process the sequence in both forward and backward directions. 
This allows the network to consider both preceding and succeeding elements for each sequence element.
The \textit{BiGRU} layer, also part of the architecture, enhances this capability by capturing information from both past and future contexts.
This is especially useful for understanding the dependencies in sequential data. 

\textit{CNN} layers are used to extract patterns from a k-size window in the input that is passed to the network.
They are creating $N$ features with a convolution operation between the text window and every distinct filter, and adding a bias term.

\textit{ReLU} introduces non-linearity to the model, allowing it to learn complex patterns.
It is computationally efficient and helps mitigate the vanishing gradient problem.

\textit{Dropout} is a regularization technique used to prevent overfitting.
By dropping out units, the network becomes less likely to overfit the training data.
This layer forces the model to learn redundant representations and improves generalization.

\textit{Max-pooling} layers reduce the dimensionality of the feature maps, which also reduces the computational load and helps make the detected features invariant to small translations.

\textit{Fully connected} layers aggregate the features learned by the previous layers and map them to the desired output dimensions.
They use a \textit{Sigmoid} activation function to introduce a non-linearity in probability distributions.

\subsubsection{BiLSTM\_CNN}
The BiLSTM\_CNN architecture integrates the advantages of both Bidirectional Long Short-Term Memory (BiLSTM) networks and Convolutional Neural Networks (CNNs). 
The BiLSTM layer processes the input embeddings bidirectionally, meaning it reads the sequence both forward and backward.
This allows the model to capture dependencies from both past and future contexts.
This is especially useful for sequential data, where the context from both directions can be important.
The hidden size and the number of layers are hyperparameters that define the capacity of the LSTM.

Dropout is applied to the input embeddings before passing them to the LSTM to prevent overfitting by randomly setting a fraction of the input units to zero at each update during training time.

After the BiLSTM layer, the output is passed through convolutional layers.
CNNs excel at extracting features and identifying local patterns in the data.
The first convolutional layer applies a set of filters to the LSTM output, followed by a ReLU activation function to introduce non-linearity. 
The max-pooling layer then reduces the size of the output while retaining the most important features.

The process is repeated with a second convolutional layer and another max-pooling layer.
This helps in extracting higher-level features and further reducing the size of the data.

The pooled output from the CNN layers is then passed through fully connected layers.
The first fully connected layer transforms the pooled output into a feature vector, which is then activated using ReLU.
The second fully connected layer maps the feature vector to the final output, which represents the predicted class probabilities.

The overall design aims to capture both sequential and spatial dependencies in the data.
The BiLSTM captures the temporal patterns, while the CNN layers focus on local features and hierarchical representations.

\subsubsection{BiGRU\_CNN}

The BiGRU\_CNN architecture, like the BiLSTM\_CNN, uses the strengths of both Bidirectional Gated Recurrent Units (BiGRUs) and Convolutional Neural Networks (CNNs).
This architecture is designed to use the sequential modeling capabilities of GRUs and the feature extraction of CNNs.
The BiGRU layer processes the input embeddings in both forward and backward directions, enabling it to capture dependencies from both past and future contexts. 

Dropout is used to prevent overfitting by randomly dropping a fraction of the input units at each update during training.
The output from the BiGRU layer is then passed through convolutional layers to extract features and detect local patterns.
The first convolutional layer applies filters to the GRU output.

Then, using a ReLU activation function, we introduce non-linearity.
The max-pooling layer keeps the most important features and decreases the size of the output.

Another convolutional layer and its associated max-pooling layer continue the process, extracting more abstract features and further reducing the dimensions of the data.

The resulting pooled output is fed into fully connected layers. The first fully connected layer transforms the pooled data into a feature vector, which is then activated using ReLU.
The second fully connected layer maps this feature vector to the final output, representing predicted class probabilities.

\subsubsection{CNN\_BiLSTM}

This architecture combines the power of CNNs to extract spatial features, as well as the advantages of BiLSTM to capture sequential dependencies.

The input embeddings are first processed by two convolutional layers, each followed by a ReLU activation function.
CNNs are good at feature extraction by applying filters over local regions of the input.
Max-pooling layers reduce the dimensions of the convolutional outputs, retaining the most relevant features.

The output from the CNN layers is then fed into a BiLSTM.
This layer processes the input sequence in both forward and backward directions, capturing temporal dependencies effectively.

The pooled output is passed through fully connected layers. 
The first fully connected layer transforms the BiLSTM output into a suitable feature representation, which is then activated by ReLU.
The second fully connected layer maps these features to the final output layer, which provides the predicted class probabilities.

\subsubsection{CNN\_BiGRU}
 This architecture passes the input into CNN layers, using the bidirectional GRUs to combine the feature extraction with sequential capturing extraction.
 
 The input embeddings are first processed through two convolutional layers, followed by ReLU activation functions.

 After each convolutional layer, there are two corresponding max-pooling layers.
 This helps in focusing on the most significant features extracted by the previous layer.

 The output is passed through a BiGRU layer.
 This layer processes the input both forward and backward, helping the model to understand the context from both past and future states.

After processing through the BiGRU layer, the pooled output is transformed by fully connected layers.

\subsubsection{\textsc{CleanNews}}

The pseudo-code for \textsc{CleanNews} is presented in Algorithm~\ref{alg:cleannews}.
\textsc{CleanNews} takes as input the dataset $X=\{ <t_i, v_i, c_i> | i=\overline{1, n} \}$, the graph $G = (V, E)$, the RNN type through the parameter $reccurent$, and whether the RNN is used as head or not through parameter $head$.
The dataset $X$ contains for each tweet a tuple $<t_i, v_i, c_i>$ $i=\overline{1, n}$ ($n=|X|$) with the following information: $t_i$ is the textual content, $v_i$ is the author of the content, i.e., a node, and $c_i$ is the class.
The graph $G=(V, E)$ maps the relationship between the users $v_i \in V$  using the edges $e_i \in E$, and it is used to embed the propagation graph.
The parameter $recurrent$ determines whether the neural network uses BiLSTMs or BiGRUs units.
The parameter $we\_model$ is used to determine which word embedding is used to encode the textual data, while the parameter $ne\_model$ determines the node embedding used to encode the propagation graph.

\begin{algorithm*}[!ht]
\small
\SetAlgoNoLine
\DontPrintSemicolon
\newcommand{\hrulealg}[0]{\vspace{1mm} \hrule \vspace{1mm}}

\SetKwInOut{Input}{Input}
\SetKwInOut{Output}{Output}
\SetKwFor{Loop}{Loop}{}{EndLoop}
\Input{dataset $X=\{ <t_i, v_i, c_i> | i=\overline{1, n} \}$ \newline
    graph $G=\{ V, E \}$ \newline
    RNN type $reccurent$   \newline
    use RNN head $head$ \newline
    word embedding model for the Text Branch $we\_model$ \newline
    node embedding model for the Propagation Branch $ne\_model$
    }
\Output{\textsc{CleanNews} model $m$}

\emph{$T \gets \emptyset$}\label{line01} \atcp{text list}
\emph{$V \gets \emptyset$}\label{line02} \atcp{nodes list}
\emph{$C \gets \emptyset$}\label{line03} \atcp{label list}

\ForEach{$<t, v, c> \in X$}{\label{line04}
    \emph{$T \gets t \cup \{t\}$}\label{line05} \atcp{populate the text list}
    \emph{$V \gets V \cup \{v\}$}\label{line06} \atcp{populate the node list}
    \emph{$C \gets C \cup \{c\}$}\label{line07} \atcp{populate the label list}
}

\emph{$P_{texts} \gets getPreprocessTexts(T)$}\label{line08} \atcp{preprocess the textual data}
\emph{$D \gets getDocument2Token(P_{texts})$}\label{line09} \atcp{get document to token matrix}
\emph{$W \gets getWordEmbeding(D, we\_model)$}\label{line10} \atcp{train the word embedding and encode tweets}
\emph{$N \gets getNodeEmbeding(G, V, ne\_model)$}\label{line11} \atcp{train the node embedding}

\emph{$Concat_{emb} \gets concatenate(W, N)$}\label{line12}
\atcp{concatenate embeddings}

\If{$reccurent = GRU$}{\label{line13}
    \emph{$RNN = BiGRU()$}\label{line14}
}
\ElseIf{$reccurent = LSTM$}{\label{line15}
    \emph{$RNN = BiLSTM()$}\label{line16}
}

\emph{$i\_layer \gets input(dims = Concat_{emb}.shape)$}\label{line17}
\atcp{deep neural network input layer}
\emph{$dnn.addLayer(i\_layer)$}\label{line18}

\If{$head = true$}{\label{line19}
    \emph{$dnn.addLayer(Dropout())$}\label{line20}
    
    \emph{$dnn.addLayer(RNN)$}\label{line21}
} 

\emph{$dnn.addLayer(CNN())$}\label{line22}

\emph{$dnn.addLayer(RelU())$}\label{line23}

\emph{$dnn.addLayer(MaxPolling())$}\label{line24}

\emph{$dnn.addLayer(CNN())$}\label{line25}

\emph{$dnn.addLayer(RelU())$}\label{line26}

\emph{$dnn.addLayer(MaxPolling())$}\label{line27}

\If{$head = false$}{\label{line28}
    \emph{$dnn.addLayer(RNN)$}\label{line29}
} 

\emph{$dnn.addLayer(FullyConnected())$}\label{line30}

\emph{$dnn.addLayer(RelU())$}\label{line31}

\emph{$o\_layer \gets dnn.addLayer(FullyConnected())$}\label{line32}
\atcp{deep neural network output layer}

\emph{$m \gets Model(input=i\_layer, output=o\_layer)$}\label{line33} \atcp{model definition}

\emph{$m.train(Concat_{emb}, C) $}\label{line34}  \atcp{model training}

\Return{$m$}\;\label{line35}
\caption{\textsc{CleanNews}}
\label{alg:cleannews}
\end{algorithm*}\DecMargin{1em}

The algorithm's workflow is as follows.
We initialize and populate the $T$, $V$, and $C$ lists where we store the text $t_i$, the nodes $v_i$, and the class $c_i$ for each entry in the dataset $X$ (Lines~\ref{line01}-\ref{line07}).
After initialization, the lists contain information as follows: 
$T=\{t_i| i=\overline{1, n}\}$, 
$V=\{v_i | i=\overline{1, n}\}$, and 
$C=\{c_i| i=\overline{1, n}\}$. 
The list $T$ is used to preprocess the textual data $P_{texts}$ (Line~\ref{line08}).
The preprocessed textual data $P_{texts}$ is used to obtain the document-to-token matrix $D$ (Line~\ref{line09}). 
The function $getWordEmbeding(D, we\_model)$ is uses to create the word embeddings $W \in \mathbb{R}^{||X|| \times t}$ ($t$ is the word embedding dimension) using the $we\_model$ parameter to determine which word embedding model to use and the document term matrix $D$ (Line~\ref{line10}). 
The function $getNodeEmbeding(G, V, ne\_model)$ uses the Graph $G$, the node list $V$, i.e., the users list, and the $ne\_model$ parameter to create the node embeddings $N \in \mathbb{R}^{||V|| \times s}$ (Line~\ref{line11}), where $s$ the dimension of the node embedding.
Finally, we concatenate the tweet and node embeddings $Concat_{emb}$ (Line~\ref{line12}) as follows: for each tweet concatenate the tweet embedding and its user's node embedding.
As our architecture is modular, any word or transformer and node embedding can be used.

Lines~\ref{line13}-~\ref{line14} determine which RNN cells to use, i.e., BiGRU or BiLSTM.
Lines~\ref{line17}-\ref{line32} define the deep neural network used for classification as presented in Figure~\ref{fig:CleanNewsDetArch}.
First, we define the input layer (Line~\ref{line17}) that accepts inputs multidimensional matrices of size equal to the dimensions of the concatenated embedding $Concat_{emb} \in \mathbb{R}^{n \times m}$, i.e., $n \times m$, where $n=||X||$ is the number of records and $m = t + s$ is dimension of the word embedding $t$ plus the dimension of the node embedding $s$. 
Second, we add the layer of the architecture (Lines~\ref{line18}-\ref{line31}) and define the output layer (Line~\ref{line32}), which uses the final Fully Connected layer. 
Third, the model $m$ is defined (Line~\ref{line33}) using as input and output layers.
Finally, the model is trained (Line~\ref{line34}) and returned (Line~\ref{line35}).

\subsection{Fake News Mitigation}

In the Mitigation stage, when a post is identified as false, a process is initiated to prevent its further propagation within the network.
This involves the implementation of a network immunization algorithm.
The process strategically targets the nodes that are directly linked to the source of the false information. 

By deploying Immunization algorithms in the Mitigation stage, the network can efficiently counterbalance the spread of false information.
We use SparseShield and NetShield algorithms to mitigate the spread of harmful content posted by malicious nodes.
As a result, both algorithms prioritize nodes that are not harmful when deciding what nodes to immunize.
To immunize the network, the harmful nodes receive a lower score for ranking (Equation~\eqref{eq:score}), thereby minimizing their effect.

\begin{equation}\label{eq:score}
    \text{score} = \text{score} \cdot 0.5\
\end{equation}

\subsubsection{SparseShield}
This algorithm minimizes the spread of influence within a graph. This helps immunize the network.

The original approach uses a priority queue to select the nodes to block, based on their impact on the graph's connectivity.

The algorithm begins by selecting the essential components: the list of nodes and their count.
It constructs an index mapping to facilitate the node-to-index and index-to-node conversions.

Using the adjacency matrix of the graph, the algorithm determines the largest eigenvalue and its corresponding eigenvector.

A priority queue is initialized to manage nodes based on their score, which reflects their influence.
The harmful nodes receive a lower score (Equation~\eqref{eq:score}) to mitigate their impact.

The algorithm selects $k$ nodes from the priority queue, representing the set of nodes to be immunized (blocked).
For each selected node, its neighbors' scores are updated in the priority queue based on their interaction with the chosen node.

\subsubsection{NetShield}
The NetShield algorithm has the same functionality as SparseShield, using a different approach.

It follows the same steps, creating an inverse mapping for node indexing.
It also uses an adjacency matrix representation of the graph to compute the eigenvalue and its corresponding eigenvector. 

A priority queue manages nodes based on their computed scores, reflecting their potential influence on the graph.
Harmful nodes get a smaller score to mitigate their impact.

The algorithm selects $k$ nodes from the priority queue.
Nodes are added to a set representing the nodes to be blocked. 
For each node, its neighbors' scores are updated in the priority queue.

The difference between the two algorithms is that NetShield uses a dense representation of the graph's adjacency matrix, while SparseShield uses a sparse representation.
This makes SparseShield more memory-efficient.

\subsubsection{Random Solver}
This immunization method provides a straightforward approach to stop the spread of harmful content.
The approach selects $k$ nodes at random to be immunized.
This method does not prioritize nodes based on their influence within the graph. 
This means that it only relies on random choices to mitigate the false information.

\subsection{Graphical User Interface}

\begin{figure}[!htbp]
  \centering
  \includegraphics[width=0.75\columnwidth]{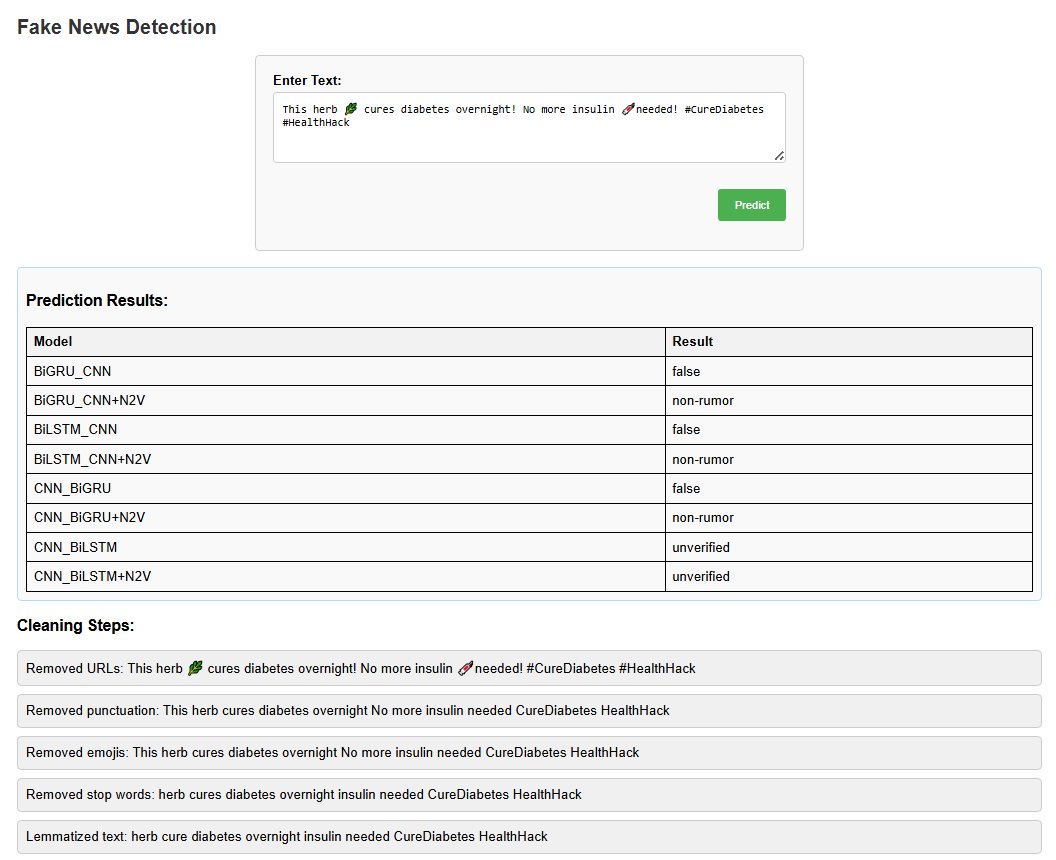}
  \caption{Fake news detection result}
  \label{fig:fnresult}
\end{figure}

\begin{figure}[!htbp]
  \centering
  \includegraphics[width=0.75\columnwidth]{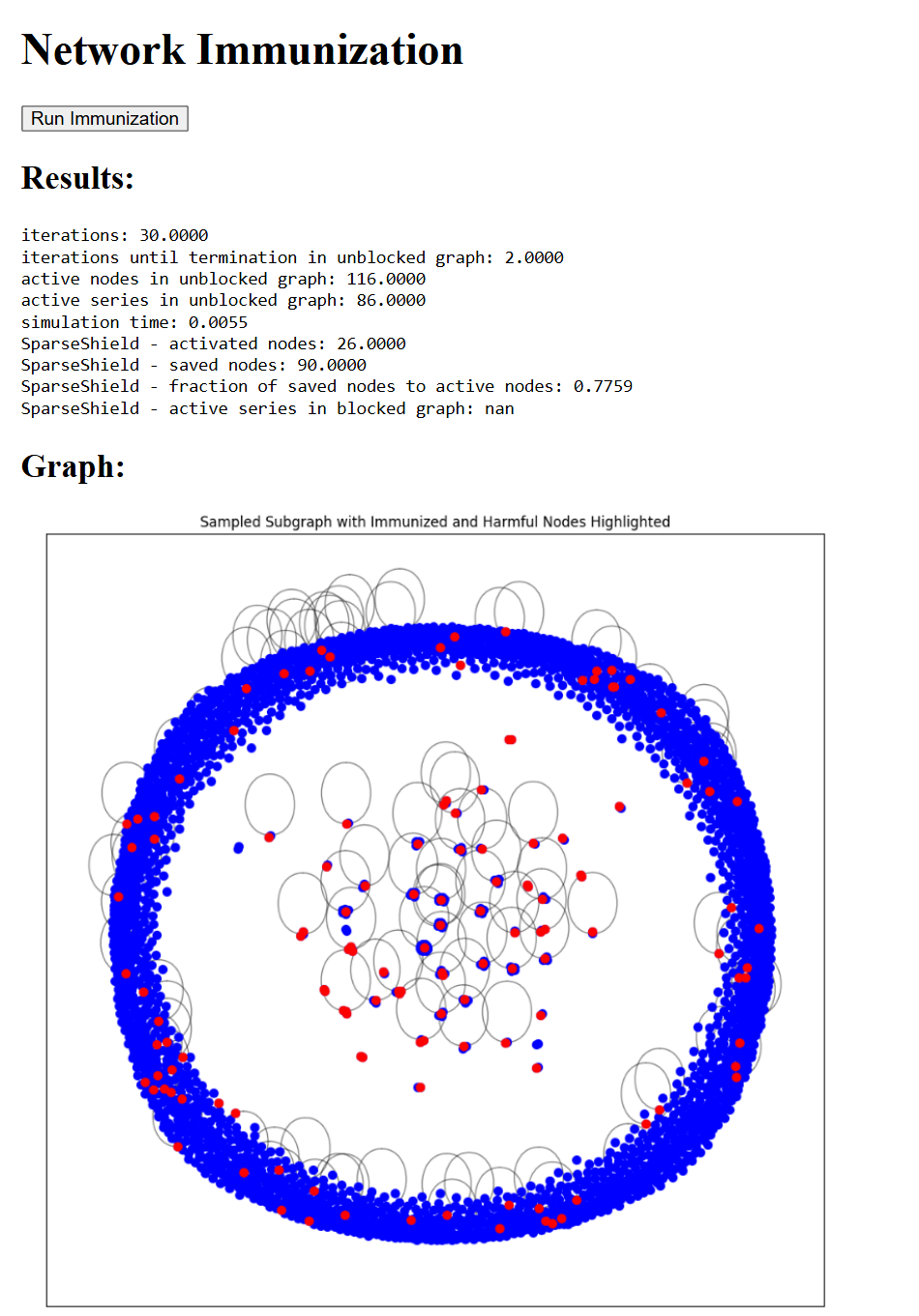}
  \caption{Fake news mitigation}
  \label{fig:fnmitigation}
\end{figure}

We built a graphical user interface for the fake news detection stage.
The user inputs the content of a post in a text box (Figure~\ref{fig:fnresult}) and clicks the \textit{Predict} button.
After that, the page displays the results of the final trained models (Table~\ref{table:finalmodels}).
Under the results, the user can see the preprocessing applied to the text.
On the second page (Figure~\ref{fig:fnmitigation}), we can see the immunization process done with SparseShield. 
First, the user enters the page and clicks \textit{Run Immunization}. 
After that, we can see the results of the immunization process and the immunized graph.

\begin{table*}[!htbp]
\centering
\caption{Fake News Detection Models}
\label{table:finalmodels}
\begin{tabular}{lcll} 
 \hline
 \textbf{Model}  & \textbf{\begin{tabular}[c]{@{}c@{}}Word\\ Embeddings\end{tabular}} & \textbf{\begin{tabular}[c]{@{}c@{}}Network\\ Embeddings\end{tabular}} & \textbf{\begin{tabular}[c]{@{}c@{}}Deep Learning\\ Architecture \end{tabular}} \\ 
 \hline
BiLSTM\_CNN  & DeBERTa & N/A      & BiLSTM\_CNN \\ 
BiGRU\_CNN &  DeBERTa & N/A      & BiGRU\_CNN \\
BiLSTM\_CNN+N2V &  DeBERTa & Node2Vec & BiLSTM\_CNN\\
BiGRU\_CNN+N2V &  DeBERTa & Node2Vec & BiGRU\_CNN\\
CNN\_BiLSTM &  DeBERTa & N/A      & CNN\_BiLSTM\\
CNN\_BiGRU &  DeBERTa & N/A      & CNN\_BiGRU\\
CNN\_BiLSTM+N2V &  DeBERTa & Node2Vec & CNN\_BiLSTM\\
CNN\_BiGRU+N2V &  DeBERTa & Node2Vec & CNN\_BiGRU\\
 \hline
\end{tabular}
\end{table*}

\section{Experimental Results}\label{sec:results}

In this section, we present the datasets, the metrics used for evaluation, and our detection and mitigation experimental results. 
The code is publicly available on GitHub at \url{https://github.com/DS4AI-UPB/CleanNews}.

\subsection{Datasets}

To determine if our models generalize well, we use two datasets for our experiments: \href{https://github.com/majingCUHK/Rumor_RvNN}{Twitter15} and \href{https://github.com/majingCUHK/Rumor_RvNN}{Twitter16} \cite{Ma2017}.
Table \ref{tab:data_stats} presents the statistics for these datasets.
\citet{Ma2017} present an in-depth analysis of \href{https://github.com/majingCUHK/Rumor_RvNN}{Twitter15} and \href{https://github.com/majingCUHK/Rumor_RvNN}{Twitter16} .

\begin{table}[!htbp]
\centering
\caption{Datasets statistics}\label{tab:data_stats}
\resizebox{1\columnwidth}{!}{%
\begin{tabular}{lrrrl}
\hline
\multicolumn{1}{c}{\textbf{Dataset}} & \multicolumn{1}{c}{\textbf{\begin{tabular}[c]{@{}c@{}}Corpus\\ Size\end{tabular}}} & \multicolumn{1}{c}{\textbf{\begin{tabular}[c]{@{}c@{}}Document\\ Size\end{tabular}}} & \multicolumn{1}{c}{\textbf{\begin{tabular}[c]{@{}c@{}}Vocabulary\\ Size\end{tabular}}} & \multicolumn{1}{c}{\textbf{\begin{tabular}[c]{@{}c@{}}Classes labels\\ (Class: encoding)\end{tabular}}}    \\
\hline
\multirow{2}{*}{Twitter15}           & \multirow{2}{*}{1\,490}                                                            & \multirow{2}{*}{52}                                                                  & \multirow{2}{*}{4\,258}                                                                & \multirow{4}{*}{\begin{tabular}[c]{@{}l@{}}true: 0\\ false: 1\\ unverified: 2\\ non-rumor: 3\end{tabular}} \\
                                     &                                                                                    &                                                                                      &                                                                                        &                                                                                                            \\
\multirow{2}{*}{Twitter16}           & \multirow{2}{*}{818}                                                               & \multirow{2}{*}{27}                                                                  & \multirow{2}{*}{2\,795}                                                                &                                                                                                            \\
                                     &                                                                                    &                                                                                      &                                                                                        &       \\ \hline                                                                                                    
\end{tabular}
}
\end{table}

Twitter15 consists of 1\,490 tweets, while Twitter16 includes 818 tweets~\cite{Ma2017}.
Each tweet is categorized into one of four classes: \textit{true}, \textit{false}, \textit{unverified}, or \textit{non-rumor}.
After preprocessing, the maximum document length in Twitter15 is 52 tokens, with a vocabulary size of 4\,258 lemmas.
In Twitter16, the maximum document length is 27 tokens, and the vocabulary comprises 2\,795 lemmas.

\subsection{Evaluation Metrics}

To evaluate the performance of each model, we measured Accuracy, Precision, Recall, and F1-Score, taking into account class imbalance~\cite{Truica2017}.
Since each dataset was a multi-class classification problem, the last three scores were calculated accordingly to reflect the performance across all classes.

\textbf{Accuracy} measures the proportion of correctly classified instances among the total number of instances: $\text{Accuracy} = \frac{\text{TP} + \text{TN}}{\text{TP} + \text{TN} + \text{FP} + \text{FN}}$.  
\textbf{Precision} is the proportion of true positive predictions among all positive predictions made by the model: 
$\text{Precision} = \frac{\text{TP}}{\text{TP} + \text{FP}}$.
\textbf{Recall} measures the ratio of actual positives that were correctly predicted by the model: $\text{Recall} = \frac{\text{TP}}{\text{TP} + \text{FN}}$.
\textbf{F1-Score} is the harmonic mean of Precision and Recall, providing a balance between the two metrics: $\text{F1-Score} = 2 \times \frac{\text{Precision} \times \text{Recall}}{\text{Precision} + \text{Recall}}$.

\subsection{Twitter15}
\label{subsec:Twitter15}

\subsubsection{Hyperparameters tuning}

To identify the best parameters for each model, we implemented a Grid Search.
This is an important technique that helps fine-tune the hyperparameters of the models.
The tested parameters were: $hidden\_size$, $num\_layers$, $dropout$, $lr$, $numb\_epochs$, $batch\_size$.
Table~\ref{tab:hyperparams} summarizes the tested values and their descriptions.

\begin{table*}[!ht]
\centering
\caption{Hyperparameter Grid and Descriptions}
\label{tab:hyperparams}
\begin{tabular}{llp{10cm}}
\hline
\textbf{Hyperparameter} & \textbf{Values Tested} & \textbf{Description} \\
\hline
\texttt{hidden\_size} & 256, 512 & Size of the hidden state in LSTM/GRU layers. Larger sizes capture complex patterns but increase overfitting risk. \\
\texttt{num\_layers} & 3, 4 & Number of stacked recurrent layers. More layers extract deeper features but raise computational cost. \\
\texttt{dropout} & 0.1, 0.2 & Regularization by randomly zeroing inputs during training to prevent overfitting. \\
\texttt{lr} (learning rate) & 0.0001, 0.0003 & Controls the step size in optimization. Lower values stabilize training; higher values speed it up. \\
\texttt{num\_epochs} & 30, 40 & Number of full passes through the training set. More epochs may improve learning but risk overfitting. \\
\texttt{batch\_size} & 24 & Number of samples per update. Smaller sizes generalize better; larger ones train faster. \\
\hline
\end{tabular}
\end{table*}

Table~\ref{table:best_param_Twitter15} presents the best parameters for different DeBERTa models used in the Fake News Detection module.
Each model is characterized by specific settings for several hyperparameters.

\begin{table*}[!htbp]
\centering
\caption{The best parameters of DeBERTa Models for Twitter15}
\label{table:best_param_Twitter15}
\begin{tabular}{lrrlrrr}
\hline
\textbf{Model} &  \textbf{hidden\_size} & \textbf{num\_layers} & \textbf{dropout} & \textbf{lr} & \textbf{num\_epochs} & \textbf{batch\_size} \\
\hline
BiLSTM\_CNN & 256 & 3 & 0.2 & 0.0003 & 30 & 24 \\
BiGRU\_CNN & 512 & 3 & 0.1 & 0.0003 & 30 & 24 \\
BiLSTM\_CNN+N2V & 256 & 4 & 0.2 & 0.0003 & 40 & 24 \\
BiGRU\_CNN+N2V & 512 & 3 & 0.1 & 0.0003 & 30 & 24 \\
CNN\_BiLSTM & 256 & 4 & 0.1 & 0.0003 & 30 & 24 \\
CNN\_BiGRU & 256 & 4 & 0.2 & 0.0003 & 40 & 24 \\
CNN\_BiLSTM+N2V & 512 & 4 & 0.2 & 0.0003 & 40 & 24 \\
CNN\_BiLSTM+N2V & 512 & 4 & 0.2 & 0.0003 & 40 & 24 \\
\hline
\end{tabular}
\end{table*} 

\subsubsection{Ablation Testing}

\begin{figure}[!ht]
  \centering
  \includegraphics[width=0.95\columnwidth]{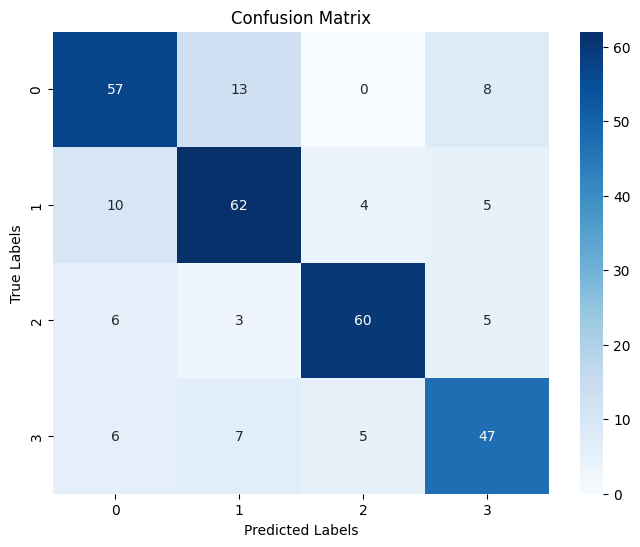}
  \caption{Confusion Matrix for CNN\_BiGRU+N2V}
  \label{fig:confmatrix8}
\end{figure}

After deciding the best parameters for each model, we perform ablation testing and evaluate the models' performance with respect to Accuracy, Precision, Recall, and F1-Score (Table~\ref{table:ablationTw15}).
The model with the best accuracy was CNN\_BiLSTM+N2V, but if we look at precision, the highest value was $0.908$ for true news, predicted by CNN\_BiLSTM.
CNN\_BiGRU predicts with the highest precision the fake news class.
The non-rumor and unverified news classes are harder to predict, as their precision scores are lower.
The confusion matrix for CNN\_BiLSTM+N2V helps visualize the model's performance better (Figure~\ref{fig:confmatrix8}).

\begin{table*}[!ht]
\centering
\caption{Ablation Study Results on Twitter15 Dataset}
\label{table:ablationTw15}
\begin{tabular}{lllrlrrr}
\hline
\textbf{Model} & \textbf{Variant} & \textbf{Embeddings} & \textbf{Accuracy} & \textbf{Class} & \textbf{Precision} & \textbf{Recall} & \textbf{F1-Score} \\
\hline

BiLSTM\_CNN & Base & DeBERTa & 0.728 & false & 0.804 & 0.577 & 0.672 \\
           &      &         &       & non-rumor & 0.649 & 0.790 & 0.707 \\
           &      &         &       & true & 0.810 & 0.865 & 0.837 \\
           &      &         &       & unverified & 0.698 & 0.677 & 0.688 \\

           & +Node2Vec & DeBERTa+Node2Vec & 0.725 & false & 0.758 & 0.641 & 0.695 \\
           &           &                  &       & non-rumor & \textbf{0.750} & 0.704 & 0.726 \\
           &           &                  &       & true & 0.727 & 0.865 & 0.790 \\
           &           &                  &       & unverified & 0.662 & 0.692 & 0.677 \\
\hline

BiGRU\_CNN  & Base & DeBERTa & 0.745 & false & 0.790 & 0.628 & 0.700 \\
           &      &         &       & non-rumor & 0.670 & 0.778 & 0.720 \\
           &      &         &       & true & 0.782 & 0.824 & 0.803 \\
           &      &         &       & unverified & 0.767 & 0.754 & 0.760 \\

           & +Node2Vec & DeBERTa+Node2Vec & 0.742 & false & 0.709 & 0.718 & 0.713 \\
           &           &                  &       & non-rumor & 0.667 & 0.765 & 0.713 \\
           &           &                  &       & true & 0.890 & 0.770 & 0.826 \\
           &           &                  &       & unverified & \textbf{0.742} & 0.708 & 0.724 \\
\hline

CNN\_BiLSTM & Base & DeBERTa & 0.695 & false & 0.833 & 0.449 & 0.583 \\
           &      &         &       & non-rumor & 0.707 & 0.824 & 0.751 \\
           &      &         &       & true & \textbf{0.908} & 0.770 & 0.832 \\
           &      &         &       & unverified & 0.495 & 0.769 & 0.602 \\

           & +Node2Vec & DeBERTa+Node2Vec & 0.688 & false & 0.830 & 0.500 & 0.624 \\
           &           &                  &       & non-rumor & 0.629 & 0.691 & 0.659 \\
           &           &                  &       & true & 0.788 & 0.851 & 0.818 \\
           &           &                  &       & unverified & 0.573 & 0.723 & 0.639 \\
\hline

CNN\_BiGRU  & Base & DeBERTa & 0.685 & false & \textbf{0.837} & 0.526 & 0.646 \\
           &      &         &       & non-rumor & 0.634 & 0.790 & 0.703 \\
           &      &         &       & true & 0.656 & 0.689 & 0.667 \\
           &      &         &       & unverified & 0.696 & 0.738 & 0.716 \\

           & +Node2Vec & DeBERTa+Node2Vec & \textbf{0.758} & false & 0.722 & 0.731 & 0.726 \\
           &           &                  &       & non-rumor & 0.730 & 0.765 & 0.747 \\
           &           &                  &       & true & 0.870 & 0.811 & \textbf{0.839} \\
           &           &                  &       & unverified & 0.723 & 0.723 & 0.723 \\
\hline
\textsc{CleanNews}    & +Node2Vec & DeBERTa+Node2Vec  & \textbf{0.758} & overall & 0.722 & 0.731 & 0.726 \\
\hline

\end{tabular}
\end{table*}

\subsubsection{Comparison}

Table~\ref{table:comparison15} presents a comparison with state-of-the-art models on the Twitter15 dataset. 
We employ the same models used by~\cite{Lu2020GCAN} and~\cite{Truica2025} in their comparison.
DTC~\cite{Castillo2011} applies decision tree learning on features derived from both user and tweet data.
SVM-TS~\cite{Ma2016detecting} relies on a support vector machine trained with features from the source tweet and the corresponding retweet sequence.
mGUR~\cite{Ma2016detecting} introduces a modified GRU network that incorporates source tweet content alongside retweet dynamics.
RFC~\cite{Kwon2017} adopts a random forest classifier utilizing information from the source tweet and user retweets.
tCNN~\cite{Yang2018} extends convolutional neural networks to learn local structures within user profile sequences and tweet content.
CRNN~\cite{Liu2018} integrates recurrent and convolutional architectures to process the source tweet information.
CSI~\cite{Ruchansky2017} combines tweet content with user credibility scores inferred from retweet behavior.
dEFEND~\cite{Shu2019} employs a co-attention mechanism to jointly model source tweets and user profiles.
GCAN and GCAN-G~\cite{Lu2020GCAN} incorporate user features into retweet propagation modeling, where GCAN leverages both recurrent and convolutional components, while GCAN-G omits the convolutional network.
SSGE~\cite{Vu2021rumor} employs graph embeddings to model information propagation patterns.
DANES~\cite{Truica2023danes} uses Mittens word embeddings, applying GRU to the textual component and a combination of GRU and CNN for the network component.
GETAE~\cite{Truica2025} leverages DeepWalk for network embeddings, BERTweet for word embeddings, and incorporates BiRNN layers. 
Our model, \textsc{CleanNews}, integrates DeBERTa for text representation and Node2Vec for network structure, using a CNN-BiGRU architecture. 
While \textsc{CleanNews} achieves competitive performance with an accuracy of 0.758 and an F1-Score of 0.726, 
GETAE outperforms all other models in binary classification, achieving the highest accuracy (0.827), precision (0.839), recall (0.827), and F1-Score (0.823). 
DANES also performs well, particularly in terms of precision (0.784), but lacks a reported F1-Score. 
Overall, \textsc{CleanNews} performs similarly or outperforms the state-of-the-art models used in our comparison.
These results highlight the effectiveness of combining advanced textual and structural representations in enhancing rumor detection performance for multi-class classification. 

\begin{table*}[!htbp]
    \centering
\caption{Comparison with sate-of-the-art models on Twitter15 (\textbf{*}GETAE does binary classification)}
\label{table:comparison15}
\begin{tabular}{llrrrr}
\hline
\textbf{Model} & \textbf{Embeddings} & \textbf{Accuracy} & \textbf{Precision} & \textbf{Recall} & \textbf{F1-Score} \\
\hline
DTC~\cite{Castillo2011}       & Linguistic Features               & 0.494    & 0.496          & 0.480       & 0.494   \\
SVM-TS~\cite{Ma2016detecting} & Social Context Features           & 0.519    & 0.519          & 0.518       & 0.519   \\
mGRU~\cite{Ma2016detecting}   & Word Embeddings                   & 0.554    & 0.514          & 0.514       & 0.510   \\
RFC~\cite{Kwon2017}           & User and Linguistic Features      & 0.538    & 0.571          & 0.530       & 0.464   \\
tCNN~\cite{Yang2018}          & Word Embeddings                   & 0.588    & 0.519          & 0.520       & 0.514   \\
CRNN~\cite{Liu2018}           & Propagation Embeddings            & 0.591    & 0.529          & 0.530       & 0.524   \\
CSI~\cite{Ruchansky2017}      & Temporal and Text Features        & 0.698    & 0.699          & 0.686       & 0.717   \\
dEFEND~\cite{Shu2019}         & Word Embeddings                   & 0.738    & 0.658          & 0.661       & 0.654   \\
GCAN-G~\cite{Lu2020GCAN}      & Propagation and Source Embeddings & 0.863    & 0.795          & 0.799       & 0.793   \\
SSGE~\cite{Vu2021rumor}       & Node2Vec & 0.600     & N/A        & N/A          & N/A      \\
SSGE~\cite{Vu2021rumor}       & DeepWalk & 0.500     & N/A        & N/A          & N/A      \\
DANES~\cite{Truica2023danes} & Mittens + Network Embeddings & 0.779    & 0.786          & 0.780       & 0.783   \\
GETAE~\cite{Truica2025}\textbf{*} & BERTweet + DeepWalk & 0.827 & 0.839 & 0.827 & 0.823 \\
\textcolor{MidnightBlue}{CleanNews} & \textcolor{MidnightBlue}{DeBERTa + Node2Vec} & \textcolor{MidnightBlue}{0.758} & \textcolor{MidnightBlue}{0.722} & \textcolor{MidnightBlue}{0.731} & \textcolor{MidnightBlue}{0.726} \\

\hline
\end{tabular}
\end{table*}

\subsubsection{Immunization}

For the immunization process, a Simulator is implemented to determine the spread of information within the network.
To better understand the result, only 5\% of the nodes (2\,697) of the graph are randomly selected, creating a new subgraph.
Using CNN\_BiGRU with DeBERTa embeddings, the nodes that spread false information are marked as harmful (357 in total), but only the nodes in the subgraph are used (17).
The algorithm simulates the spread of influence/information through the network (graph). 
For all three algorithms, only 5\% of the nodes are immunized (134).
Table~\ref{tab:simulation_metrics} presents the evaluation metrics for our simulation.

\begin{table*}[!htbp]
\centering
\caption{Simulation metrics}
\label{tab:simulation_metrics}
\begin{tabular}{lp{15cm}}
\hline
\textbf{Metric} & \textbf{Description} \\
\hline
\texttt{Graph} & The simulation was conducted on a normal graph, unblocked, and also on a blocked Graph (using the immunization algorithm). \\
\texttt{Active Nodes} & This metric shows how many nodes become active by the end of the simulation. \\
\texttt{Saved Nodes} & This metric indicates the number of nodes that are saved from becoming active due to the immunization algorithm. \\
\texttt{Active Series} & This metric tracks how many nodes become active during each step of the simulation. \\
\hline
\end{tabular}
\end{table*}

\paragraph{SparseShield}

\begin{figure}[!ht]
  \centering
  \includegraphics[width=1\columnwidth]{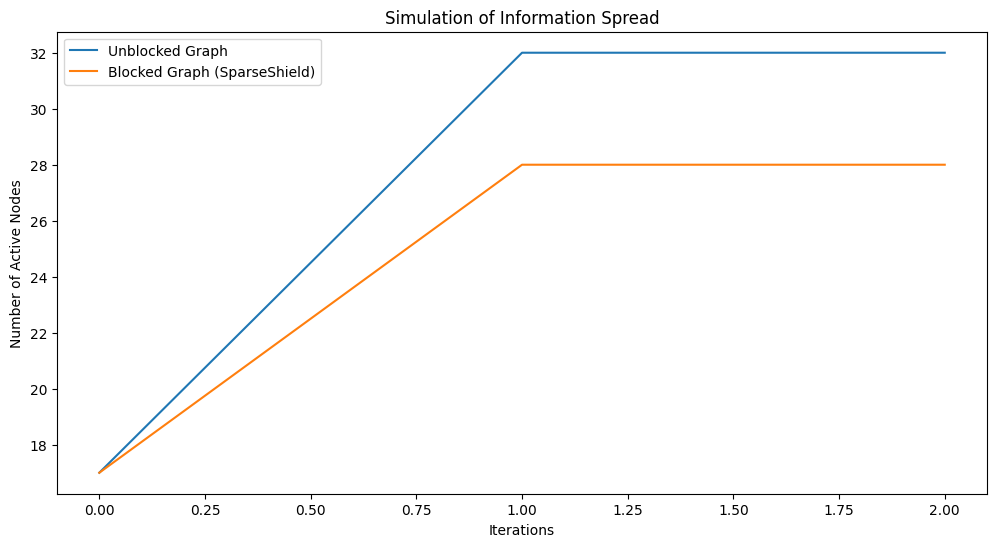}
  \caption{Simulation of Information Spread using SparseShield for Twitter15}
  \label{fig:sparseshieldspread}
\end{figure}

For SparseShield (Table~\ref{table:sparseshield15}), on average, the influence spread ends after two iterations.
Without a mitigation strategy, on average, 34 nodes become active by the end of the simulation.
The active series shows that 17 nodes are active initially, but by the second iteration, 34 nodes are active.
With SparseShield as a blocking method, 18 nodes become active and 14 nodes are saved (Figure~\ref{fig:sparseshieldspread}).

\begin{table}[!htbp]
\centering
\caption{Results for SparseShield in Twitter15}
\label{table:sparseshield15}
\begin{tabular}{lrrc} 
 \hline
 \textbf{Graph}  & \textbf{\begin{tabular}[c]{@{}c@{}}Activated\\ Nodes\end{tabular}} & \textbf{\begin{tabular}[c]{@{}c@{}}Saved\\ Nodes\end{tabular}} & \textbf{\begin{tabular}[c]{@{}c@{}}Active\\ Series\end{tabular}}\\ \hline
 Unblocked  & 32 & 0 & [17, 32, 32]\\ 
 Blocked  & 18 & 14 & [17, 28, 28]\\ 
 \hline
\end{tabular}
\end{table}
 
\begin{figure}[!htbp]
  \centering
  \includegraphics[width=0.8\columnwidth]{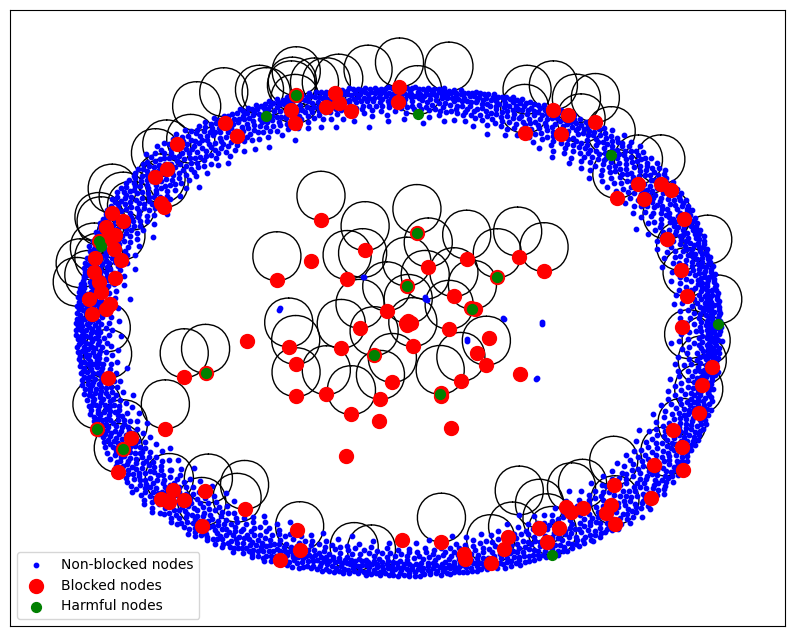}
  \caption{Nodes after SparseShield for Twitter15}
  \label{fig:sparseshieldnodes}
\end{figure}

We can also see the initial graph, with the blocked nodes and the harmful nodes (Figure~\ref{fig:sparseshieldnodes}).

\paragraph{Random Solver}

\begin{figure}[!htbp]
  \centering
  \includegraphics[width=0.8\columnwidth]{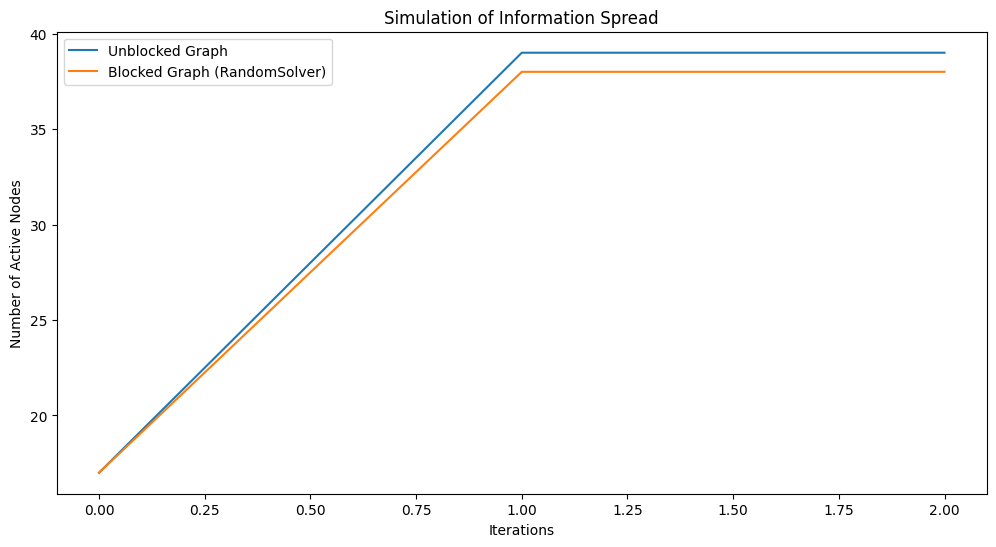}
  \caption{Simulation of Information Spread using Random Solver for Twitter15}
  \label{fig:randomspread}
\end{figure}

 For Random Solver (Table~\ref{table:randomsolver15}), on average, the influence spread ends after two iterations as well. Without mitigation, on average, 39 nodes become active by the end of the simulation.
 The active series shows that 17 nodes are active initially, and after another iteration, 39 nodes.
 With a Random blocking method, only 1 node is saved (Figure~\ref{fig:randomspread}).

\begin{table}[!htbp]
\centering
\caption{Results for Random Solver in Twitter15}
\label{table:randomsolver15}
\begin{tabular}{lrrc} 
 \hline
 \textbf{Graph}  & \textbf{\begin{tabular}[c]{@{}c@{}}Activated\\ Nodes\end{tabular}} & \textbf{\begin{tabular}[c]{@{}c@{}}Saved\\ Nodes\end{tabular}} & \textbf{\begin{tabular}[c]{@{}c@{}}Active\\ Series\end{tabular}}\\ \hline
 Unblocked  & 39 & 0 & [17, 39, 39]\\ 
 Blocked  & 38 & 1 & [17, 38, 38]\\ 
 \hline
\end{tabular}
\end{table}

\begin{figure}[!htbp]
  \centering
  \includegraphics[width=0.8\columnwidth]{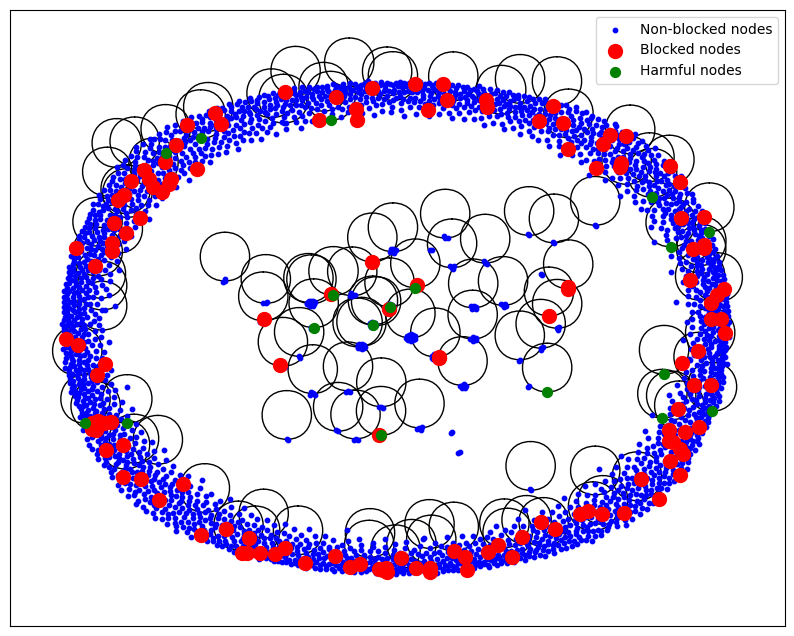}
  \caption{Nodes after Random Solver for Twitter15}
  \label{fig:randomnodes}
\end{figure}

Figure~\ref{fig:randomnodes} shows that the blocked nodes are not necessarily close to the harmful.

\paragraph{NetShield}

\begin{figure}[!htbp]
  \centering
  \includegraphics[width=0.8\columnwidth]{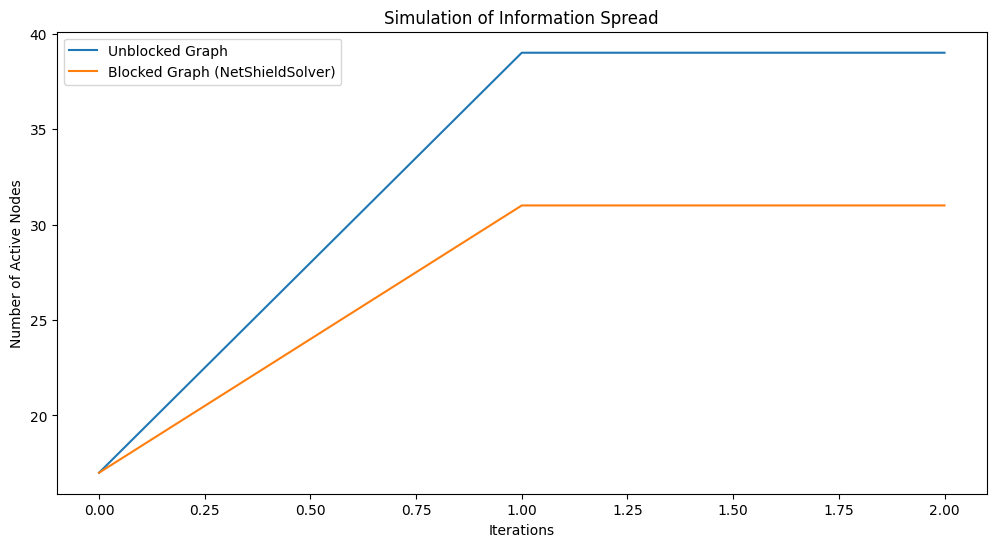}
  \caption{Simulation of Information Spread using NetShield for Twitter15}
  \label{fig:netshieldpread}
\end{figure}

For NetShield (Table~\ref{table:netshield15}), without mitigation, on average, 39 nodes become active by the end of the simulation.
Again, the active series shows 17 nodes active for the first iteration, and 39 nodes after.
With a NetShield blocking strategy, 11 nodes are saved (Figure~\ref{fig:netshieldpread}).

\begin{table}[!ht]
\centering
\caption{Results for NetShield in Twitter15}
\label{table:netshield15}
\begin{tabular}{lrrc} 
 \hline
 \textbf{Graph}  & \textbf{\begin{tabular}[c]{@{}c@{}}Activated\\ Nodes\end{tabular}} & \textbf{\begin{tabular}[c]{@{}c@{}}Saved\\ Nodes\end{tabular}} & \textbf{\begin{tabular}[c]{@{}c@{}}Active\\ Series\end{tabular}}\\ \hline
 Unblocked  & 39 & 0 & [17, 39, 39]\\ 
 Blocked  & 28 & 11 & [17, 31, 31]\\ 
 \hline
\end{tabular}
\end{table}

\begin{figure}[!htbp]
  \centering
  \includegraphics[width=0.8\columnwidth]{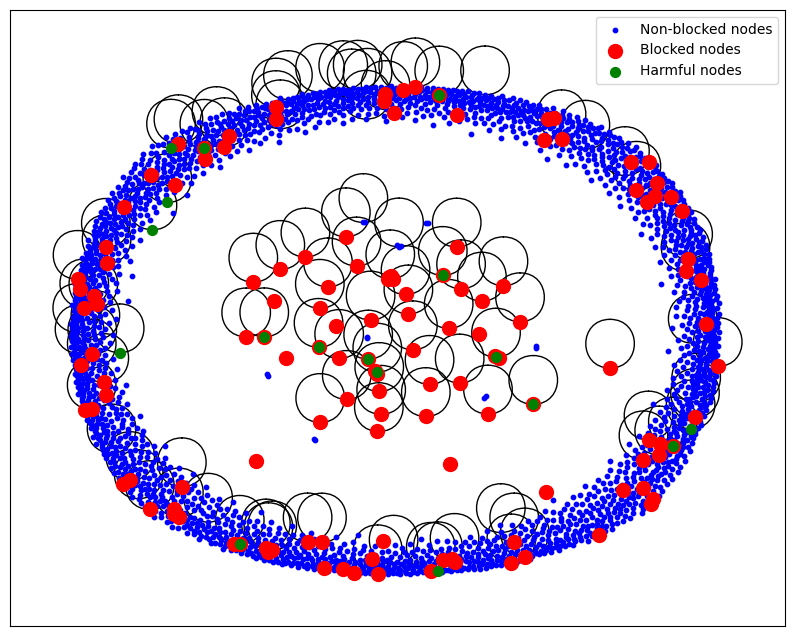}
  \caption{Nodes after NetShield for Twitter15}
  \label{fig:netshieldNodes}
\end{figure}

Unlike the previous approach, the blocked nodes are close to the harmful nodes this time (Figure~\ref{fig:netshieldNodes}).

\subsection{Twitter16}
\subsubsection{Hyperparameters tuning}
For this dataset, we used the same Grid Search parameters as for the Twitter15 dataset (Table~\ref{tab:hyperparams}).
As a result, we identified the best parameters for each model in Table~\ref{table:bestparams16}.
Each parameter has the same function as it did in the previous dataset, as explained in Section~\ref{subsec:Twitter15}.

\begin{table*}[!htbp]
\centering
\caption{The best parameters of DeBERTa Models for Twitter16}
\label{table:bestparams16}
\begin{tabular}{lrrrlrr}
\hline
\textbf{Model} & \textbf{hidden\_size} & \textbf{num\_layers} & \textbf{dropout} & \textbf{lr} & \textbf{num\_epochs} & \textbf{batch\_size} \\
\hline
BiLSTM\_CNN & 256 & 4 & 0.1 & 0.0001 & 30 & 24 \\
BiGRU\_CNN & 512 & 3 & 0.1 & 0.0003 & 30 & 24 \\
BiLSTM\_CNN+N2V & 256 & 3 & 0.1 & 0.0003 & 30 & 24 \\
BiGRU\_CNN+N2V & 256 & 4 & 0.1 & 0.0003 & 40 & 24 \\
CNN\_BiLSTM & 256 & 3 & 0.1 & 0.0003 & 40 & 24 \\
CNN\_BiGRU & 256 & 4 & 0.1 & 0.0003 & 40 & 24 \\
CNN\_BiLSTM+N2V & 256 & 3 & 0.2 & 0.0003 & 40 & 24 \\
CNN\_BiGRU+N2V & 256 & 4 & 0.1 & 0.0003 & 30 & 24 \\
\hline
\end{tabular}
\end{table*}

\subsubsection{Ablation Testing}
\begin{figure}[!htbp]
  \centering
  \includegraphics[width=0.95\columnwidth]{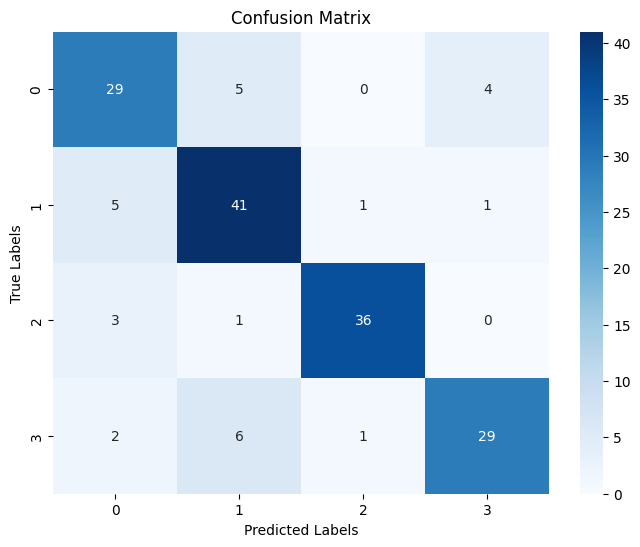}
  \caption{Confusion Matrix for BiLSTM\_CNN}
  \label{fig:confmatrix1}
\end{figure}

Table~\ref{table:resultsTw16} presents the ablation test evaluation metrics.
For this set of experiments, BiLSTM\_CNN with DeBERTa embeddings scored the highest accuracy, followed by CNN\_BiLSTM and CNN\_BiGRU, both with Node2Vec embeddings added.
We observe that the scores are significantly higher for this dataset.
This difference is best highlighted by the CNN\_BiLSTM and CNN\_BiGRU  models, where the accuracy increased by 10\%. 
The confusion matrix presents the results for the BiLSTM\_CNN model (Figure~\ref{fig:confmatrix1}).
When looking at the other scores, CNN\_BiGRU predicts false news with the highest precision.
Furthermore, the models have a hard time predicting non-rumor and unverified tweets.

\begin{table*}[!ht]
\centering
\caption{Ablation Study Results on Twitter16 Dataset}
\label{table:resultsTw16}
\begin{tabular}{lllrlrrr}
\hline
\textbf{Model} & \textbf{Variant} & \textbf{Embeddings} & \textbf{Accuracy} & \textbf{Class} & \textbf{Precision} & \textbf{Recall} & \textbf{F1-Score} \\
\hline

BiLSTM\_CNN & Base & DeBERTa & \textbf{0.823} & false & 0.744 & 0.763 & 0.753 \\
            &      &         &                & non-rumor & 0.774 & 0.854 & 0.811 \\
            &      &         &                & true & 0.947 & 0.900 & 0.923 \\
            &      &         &                & unverified & 0.853 & 0.763 & 0.806 \\

            & +Node2Vec & DeBERTa+Node2Vec & 0.793 & false & 0.833 & 0.658 & 0.735 \\
            &           &                  &       & non-rumor & 0.712 & 0.875 & 0.785 \\
            &           &                  &       & true & 0.941 & 0.800 & 0.865 \\
            &           &                  &       & unverified & 0.756 & 0.816 & 0.785 \\
\hline

BiGRU\_CNN & Base & DeBERTa & 0.768 & false & 0.950 & 0.500 & 0.655 \\
           &      &         &       & non-rumor & 0.638 & 0.917 & 0.752 \\
           &      &         &       & true & 0.881 & 0.925 & 0.902 \\
           &      &         &       & unverified & 0.788 & 0.684 & 0.732 \\

           & +Node2Vec & DeBERTa+Node2Vec & 0.781 & false & 0.730 & 0.711 & 0.720 \\
           &           &                  &       & non-rumor & 0.765 & 0.813 & 0.788 \\
           &           &                  &       & true & 0.846 & 0.825 & 0.835 \\
           &           &                  &       & unverified & 0.784 & 0.763 & 0.773 \\
\hline

CNN\_BiLSTM & Base & DeBERTa & 0.805 & false & 0.652 & 0.789 & 0.714 \\
            &      &         &       & non-rumor & 0.755 & 0.771 & 0.763 \\
            &      &         &       & true & \textbf{0.974} & 0.925 & 0.949 \\
            &      &         &       & unverified & \textbf{0.903} & 0.737 & 0.812 \\

            & +Node2Vec & DeBERTa+Node2Vec & 0.817 & false & 0.644 & 0.763 & 0.699 \\
            &           &                  &       & non-rumor & \textbf{0.796} & 0.813 & 0.804 \\
            &           &                  &       & true & 0.942 & 0.825 & 0.880 \\
            &           &                  &       & unverified & 0.943 & 0.868 & 0.904 \\
\hline

CNN\_BiGRU & Base & DeBERTa & 0.805 & false & \textbf{0.967} & 0.763 & 0.853 \\
           &      &         &       & non-rumor & 0.686 & 0.958 & 0.800 \\
           &      &         &       & true & 0.944 & 0.850 & 0.895 \\
           &      &         &       & unverified & 0.742 & 0.605 & 0.667 \\

           & +Node2Vec & DeBERTa+Node2Vec & 0.811 & false & 0.674 & 0.763 & 0.716 \\
           &           &                  &       & non-rumor & 0.755 & 0.833 & 0.792 \\
           &           &                  &       & true & \textbf{0.974} & 0.950 & 0.962 \\
           &           &                  &       & unverified & 0.897 & 0.684 & 0.776 \\
\hline
\textsc{CleanNews}    & +Node2Vec & DeBERTa+Node2Vec  & \textbf{0.823} & overall & 0.829 & 0.820 & 0.823 \\
\hline
\end{tabular}
\end{table*}

\subsubsection{Comparison}

\begin{table*}[!ht]
    \centering
\caption{Comparison with sate-of-the-art models on Twitter16 (\textbf{*}GETAE does binary classification)}
\label{table:comparison16}
\begin{tabular}{llrrrr}
\hline
\textbf{Model} & \textbf{Embeddings} & \textbf{Accuracy} & \textbf{Precision} & \textbf{Recall} & \textbf{F1-Score} \\
\hline
DTC~\cite{Castillo2011}       & Linguistic Features               & 0.561    & 0.575          & 0.536       & 0.561  \\ 
SVM-TS~\cite{Ma2016detecting} & Social Context Features           & 0.693    & 0.692          & 0.691       & 0.691  \\ 
mGRU~\cite{Ma2016detecting}   & Word Embeddings                   & 0.661    & 0.560          & 0.561       & 0.556  \\ 
RFC~\cite{Kwon2017}           & User and Linguistic Features      & 0.662    & 0.731          & 0.658       & 0.627  \\ 
tCNN~\cite{Yang2018}          & Word Embeddings                   & 0.737    & 0.624          & 0.626       & 0.620  \\ 
CRNN~\cite{Liu2018}           & Propagation Embeddings            & 0.757    & 0.641          & 0.643       & 0.636  \\ 
CSI~\cite{Ruchansky2017}      & Temporal and Text Features        & 0.661    & 0.632          & 0.630       & 0.630  \\ 
dEFEND~\cite{Shu2019}         & Word Embeddings                   & 0.701    & 0.636          & 0.638       & 0.631  \\ 
GCAN-G~\cite{Lu2020GCAN}      & Propagation and Source Embeddings & 0.793    & 0.678          & 0.680       & 0.675  \\ 
SSGE~\cite{Vu2021rumor}       & Node2Vec & 0.600     & N/A             & N/A          & N/A     \\
SSGE~\cite{Vu2021rumor}       & DeepWalk & 0.500     & N/A             & N/A          & N/A     \\
DANES~\cite{Truica2023danes} & FastText + Network Embeddings & 0.779 & 0.785 & 0.779 & 0.782 \\
GETAE~\cite{Truica2025}\textbf{*} & BERT + Node2Vec & 0.895 & 0.897 & 0.901 & 0.896 \\ 
\textcolor{MidnightBlue}{\textsc{CleanNews}} & \textcolor{MidnightBlue}{DeBERTa + Node2Vec} & \textcolor{MidnightBlue}{0.823} & \textcolor{MidnightBlue}{0.829} & \textcolor{MidnightBlue}{0.820} & \textcolor{MidnightBlue}{0.823} \\
\hline
\end{tabular}
\end{table*}

Table~\ref{table:comparison16} presents a comparison with state-of-the-art models on the Twitter16 dataset using the same models as in the previous comparison.
GETAE combines Node2Vec for network embeddings, BERT for word embeddings, and RNN layers for sequence modeling. 
DANES incorporates FastText word embeddings, using a hybrid LSTM-CNN architecture for the network and text component. 
Our model, \textsc{CleanNews}, integrates DeBERTa for capturing contextual word representations in a BiLSTM\_CNN architecture.
In terms of performance, \textsc{CleanNews} achieves strong results, with an accuracy of 0.823 and an F1-Score of 0.823, demonstrating a solid balance between precision (0.829) and recall (0.820). 
DANES performs slightly lower, with an accuracy of 0.779 and F1-Score of 0.782. 
GETAE again leads the comparison for binary classification, achieving the highest metrics across the board, including accuracy (0.895), precision (0.897), recall (0.901), and F1-Score (0.896). 
In line with the Twitter15 comparison, \textsc{CleanNews} performs similarly or outperforms the state-of-the-art models used in our comparison.
These results confirm the robustness of binary classification with GETAE on this dataset, while also highlighting that \textsc{CleanNews} remains competitive, especially considering its simpler architecture and that it performs multi-class classification as DANES.

\subsubsection{Immunization}

CNN\_BiGRU model was used to mark the nodes that spread false information.
The same workflow from the previous dataset was reproduced, using only 5\% of the nodes (1\,327), randomly selected. 
Among these nodes, 8 are harmful, and 66 are blocked during the immunization process.

\paragraph{SparseShield}

\begin{table}[!htbp]
\centering
\caption{Results for SparseShield in Twitter16}
\label{table:sparseshield16}
\begin{tabular}{lrrc} 
 \hline
 \textbf{Graph}  & \textbf{\begin{tabular}[c]{@{}c@{}}Activated\\ Nodes\end{tabular}} & \textbf{\begin{tabular}[c]{@{}c@{}}Saved\\ Nodes\end{tabular}} & \textbf{\begin{tabular}[c]{@{}c@{}}Active\\ Series\end{tabular}}\\
 \hline
 Unblocked  & 23 & 0 & [8, 32, 32]\\ 
 Blocked  & 11 & 12 & [8, 12, 12]\\ 
 \hline
\end{tabular}
\end{table}

\begin{figure}[!ht]
  \centering
  \includegraphics[width=0.8\columnwidth]{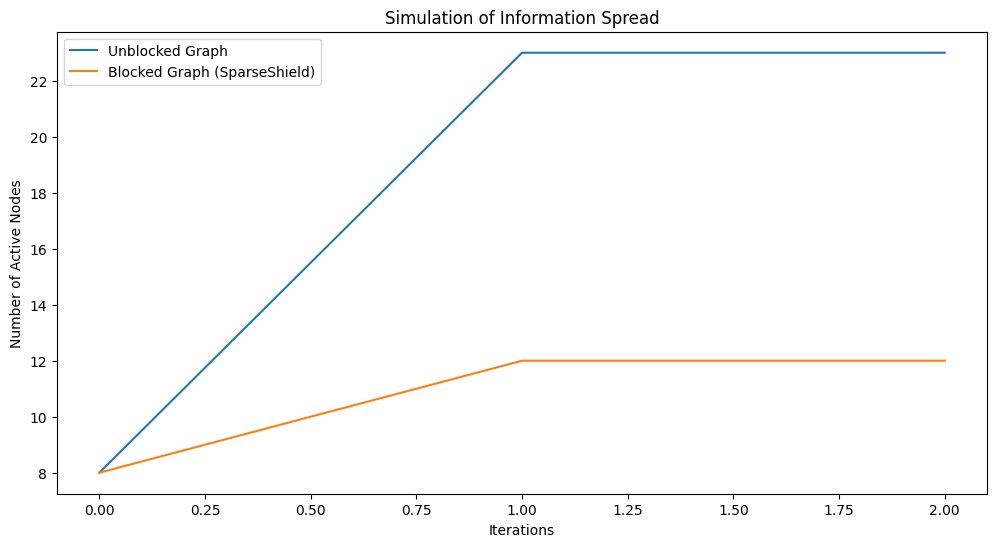}
  \caption{Simulation of Information Spread using SparseShield for Twitter16}
  \label{fig:sparseshieldspread2}
\end{figure}

During the Simulation, SparseShield saves 12 nodes.
On average, without a blocking strategy, 23 nodes are active, i.e., spreading information (Table~\ref{table:sparseshield16}).
The active series shows that 8 nodes are active initially, and by the second iteration, 32.
With SparseShield, only 12 nodes are active after the second iteration (Figure~\ref{fig:sparseshieldspread2}).

\begin{figure}[!htbp]
  \centering
  \includegraphics[width=0.8\columnwidth]{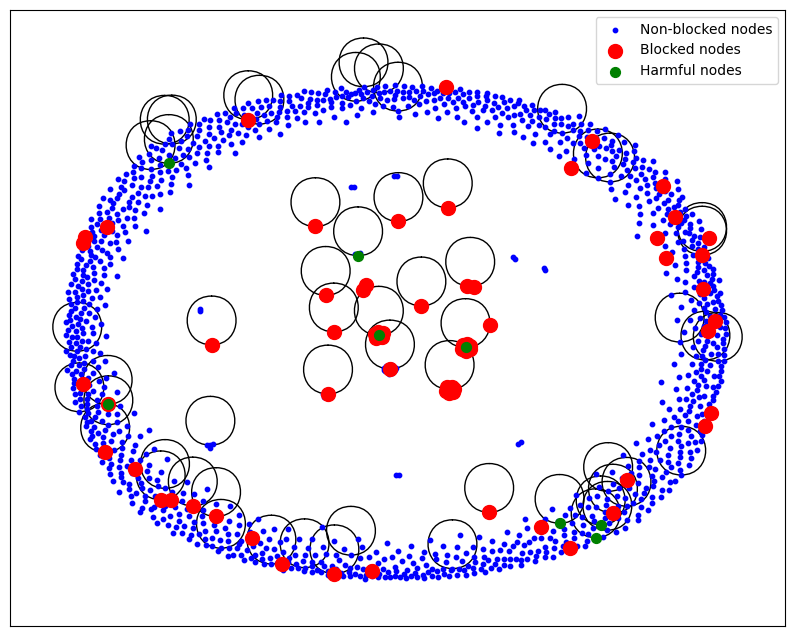}
  \caption{Nodes after SparseShield for Twitter16}
  \label{fig:sparseshieldnodes2}
\end{figure}

Figure~\ref{fig:sparseshieldnodes2} shows the immunization strategy results.

\paragraph{Random Solver}

\begin{figure}[!htbp]
  \centering
  \includegraphics[width=0.8\columnwidth]{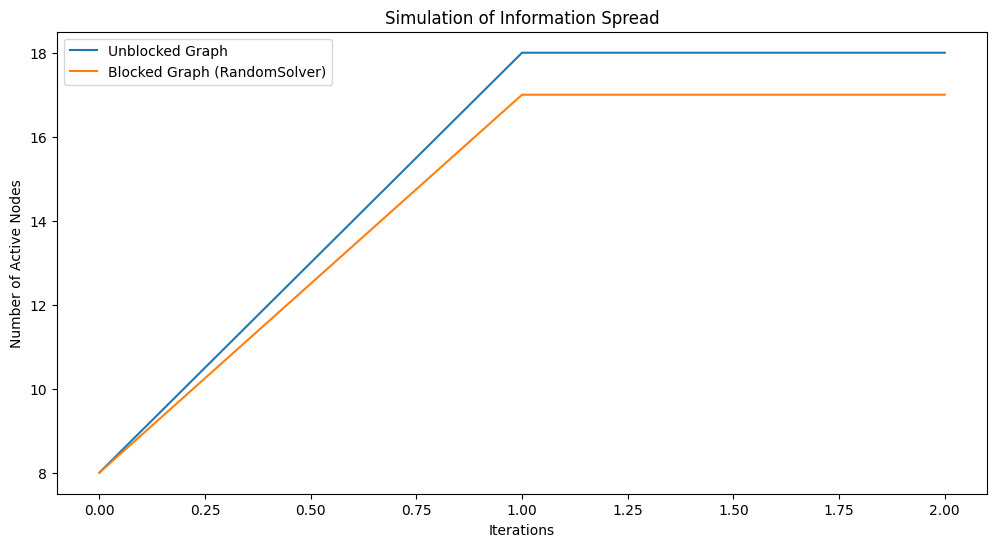}
  \caption{Simulation of Information Spread using Random Solver for Twitter16}
  \label{fig:randomspread2}
\end{figure}

Without a mitigation strategy, on average, 18 nodes are active by the end of the simulation, 8 nodes are active initially, and after another iteration, 18 nodes.
With a Random blocking strategy, we obtain the same result as before, only 1 saved node (Figure~\ref{fig:randomspread2} and Table~\ref{table:randomsolver16}).

\begin{table}[!htbp]
\centering
\caption{Results for Random Solver in Twitter16}
\label{table:randomsolver16}
\begin{tabular}{lrrc} 
 \hline
 \textbf{Graph}  & \textbf{\begin{tabular}[c]{@{}c@{}}Activated\\ Nodes\end{tabular}} & \textbf{\begin{tabular}[c]{@{}c@{}}Saved\\ Nodes\end{tabular}} & \textbf{\begin{tabular}[c]{@{}c@{}}Active\\ Series\end{tabular}}\\ \hline
 Unblocked  & 18 & 0 & [8, 18, 18]\\ 
 Blocked  & 17 & 1 & [8, 17, 17]\\ 
 \hline
\end{tabular}
\end{table}

\begin{figure}[!ht]
  \centering
  \includegraphics[width=0.8\columnwidth]{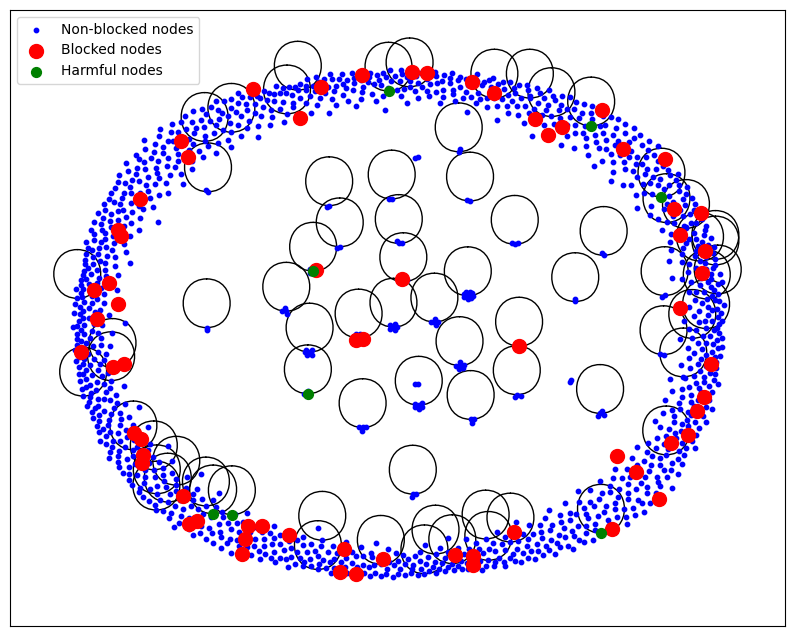}
  \caption{Nodes after Random Solver for Twitter15}
  \label{fig:randomnodes2}
\end{figure}

Figure~\ref{fig:randomnodes2} presents the results of this strategy.
We observe that the blocked nodes are not necessarily close to the harmful nodes.

\paragraph{NetShield}

NetShield saves 5 nodes, reducing the active nodes on the second iteration from 15 to 11 (Figure~\ref{fig:netshieldpread} and Table~\ref{table:netshield16}).

\begin{table}[!htbp]
\centering
\caption{Results for NetShield in Twitter16}
\label{table:netshield16}
\begin{tabular}{lrrc} 
 \hline
 \textbf{Graph}  & \textbf{\begin{tabular}[c]{@{}c@{}}Activated\\ Nodes\end{tabular}} & \textbf{\begin{tabular}[c]{@{}c@{}}Saved\\ Nodes\end{tabular}} & \textbf{\begin{tabular}[c]{@{}c@{}}Active\\ Series\end{tabular}}\\ \hline
 Unblocked  & 15 & 0 & [8, 15, 15]\\ 
 Blocked  & 10 & 5 & [8, 11, 11]\\ 
 \hline
\end{tabular}
\end{table}

\begin{figure}[!htbp]
  \centering
  \includegraphics[width=0.8\columnwidth]{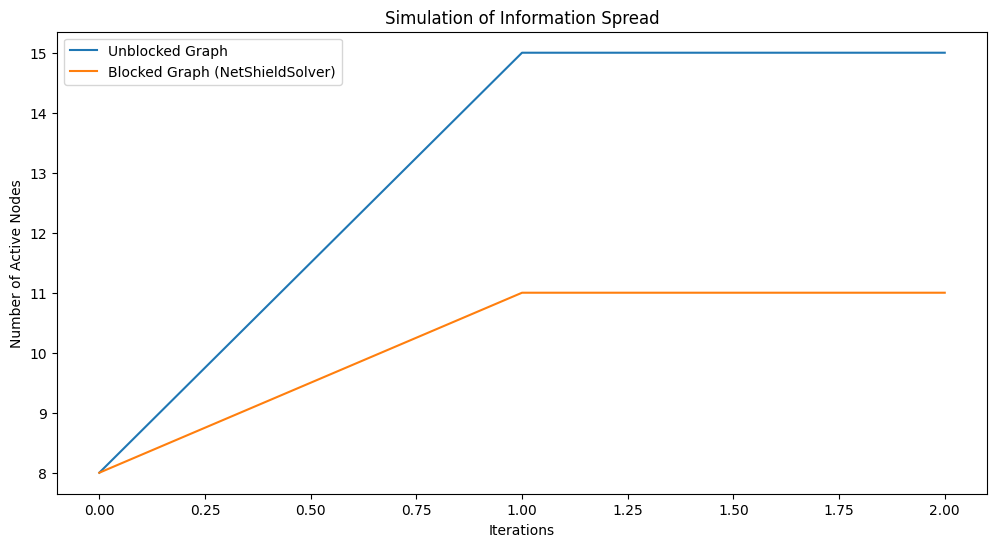}
  \caption{Simulation of Information Spread using NetShield Solver for Twitter16}
  \label{fig:netshieldpread2}
\end{figure}

\begin{figure}[!ht]
  \centering
  \includegraphics[width=0.8\columnwidth]{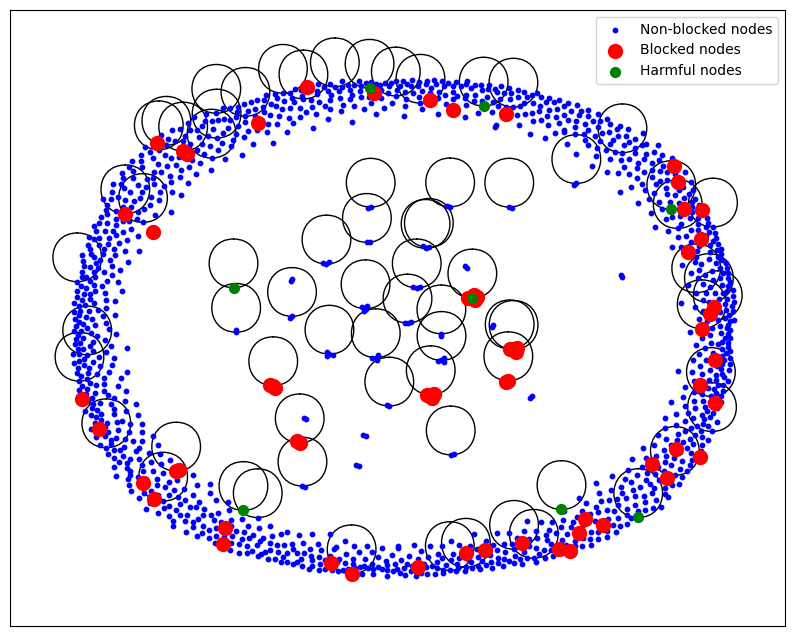}
  \caption{Nodes after NetShield for Twitter16}
  \label{fig:netshieldNodes2}
\end{figure}

This approach is definitely better than a random mitigation, but compared to SparseShield, it is weaker (Figure~\ref{fig:netshieldNodes2}).

\section{Discussions and Limitations}~\label{sec:discussion}

In this section, the results are highlighted and discussed.

\subsection{Preprocessing Module}
During this step, we experimented with classic preprocessing, aggressive preprocessing, soft preprocessing, and no preprocessing.
The results presented below are for CNN\_BiGRU, which uses DeBERTa and Node2Vec embeddings (see Table~\ref{table:diffpreproc}).
For soft preprocessing, we only eliminated emojis and punctuation marks.
The classic preprocessing was used in previous experiments.
The aggressive preprocessing included all steps from the classic preprocessing, with the addition of converting to lowercase, removing numbers, expanding contractions, removing extra whitespace, and correcting spelling.

\begin{table}[!htpb]
\centering
\caption{Model Accuracy with different types of Preprocessing}
\label{table:diffpreproc}
\begin{tabular}{lr} 
 \hline
 \textbf{Preprocessing} & \textbf{Model Accuracy} \\ 
 \hline
 No Text Clean & 0.748\\
 Soft  & 0.762\\ 
 Classic  & 0.758\\ 
 Aggressive & 0.758\\
 \hline
\end{tabular}
\end{table}

Considering the difference, a future approach would be to analyze the best strategy for text cleaning and preprocessing. 
Therefore, the best accuracy can be achieved, as well as other evaluation metrics such as precision, recall, and F1-Score.
Using this strategy, we can systematically identify which preprocessing steps contribute the most to model performance.

\subsection{Fake News Detection}

During the experiments, we used only DeBERTa embeddings for the text data.
A future approach would be to test other pre-trained models, fine-tune large language models, or use other word embeddings.
We made a first step in this direction by testing the previously implemented models on the Twitter16 dataset using GloVe embeddings \cite{Pennington2014}.
Because the results are not as satisfying as the DeBERTa results, we only use the GloVe embeddings for the eight models with the best parameters resulting from the grid search (see Table~\ref{table:best_param_Twitter15} for parameters).
Table~\ref{table:glovevsdeb} compares the two embeddings with respect to accuracy.

\begin{table}[!htbp]
\centering
\caption{Comparison of GloVe and DeBERTa Embeddings}
\label{table:glovevsdeb}
\resizebox{\columnwidth}{!}{
\begin{tabular}{lllr}
\hline
\textbf{Model} & \textbf{Variant} & \textbf{Embeddings} & \textbf{Accuracy} \\
\hline

\multirow{4}{*}{BiLSTM\_CNN} 
  & GloVe & GloVe & 0.470 \\
  &       & GloVe + Node2Vec & 0.527 \\
  & DeBERTa & DeBERTa & \textbf{0.823} \\
  &         & DeBERTa + Node2Vec & 0.793 \\
\hline

\multirow{4}{*}{BiGRU\_CNN} 
  & GloVe & GloVe & 0.651 \\
  &       & GloVe + Node2Vec & 0.553 \\
  & DeBERTa & DeBERTa & 0.768 \\
  &         & DeBERTa + Node2Vec & 0.781 \\
\hline

\multirow{4}{*}{CNN\_BiLSTM} 
  & GloVe & GloVe & 0.513 \\
  &       & GloVe + Node2Vec & 0.487 \\
  & DeBERTa & DeBERTa & 0.805 \\
  &         & DeBERTa + Node2Vec & 0.817 \\
\hline

\multirow{4}{*}{CNN\_BiGRU} 
  & GloVe & GloVe & 0.443 \\
  &       & GloVe + Node2Vec & 0.611 \\
  & DeBERTa & DeBERTa & 0.805 \\
  &         & DeBERTa + Node2Vec & 0.811 \\
\hline

\end{tabular}
}
\end{table}

Using ablation testing (Table~\ref{table:gloveRes}), we observe that DeBERTa outperforms GloVe on all architectures.
In future work, it would be beneficial to explore other state-of-the-art embeddings and language models, e.g., BERT, RoBERTa,  GPT-based, etc., to further improve the performance of the models.

\begin{table*}[!ht]
\centering
\caption{Results on Twitter16 with GloVe Embeddings}
\label{table:gloveRes}
\begin{tabular}{lllrlrrr}
\hline
\textbf{Model} & \textbf{Variant} & \textbf{Embeddings} & \textbf{Accuracy} & \textbf{Class} & \textbf{Precision} & \textbf{Recall} & \textbf{F1-Score} \\
\hline

BiLSTM\_CNN & Base & GloVe & 0.470 & false & 0.528 & 0.358 & 0.427 \\
           &      &       &       & non-rumor & 0.403 & 0.333 & 0.365 \\
           &      &       &       & true & 0.461 & 0.757 & 0.574 \\
           &      &       &       & unverified & 0.368 & 0.323 & 0.344 \\

           & +Node2Vec & GloVe+Node2Vec & 0.527 & false & 0.475 & 0.487 & 0.481 \\
           &           &                &       & non-rumor & 0.618 & 0.419 & 0.500 \\
           &           &                &       & true & 0.600 & 0.811 & 0.689 \\
           &           &                &       & unverified & 0.397 & 0.384 & 0.391 \\
\hline

BiGRU\_CNN & Base & GloVe & \textbf{0.651} & false & 0.789 & 0.577 & 0.667 \\
           &      &       &                & non-rumor & 0.600 & 0.629 & 0.614 \\
           &      &       &                & true & 0.667 & 0.865 & 0.793 \\
           &      &       &                & unverified & 0.577 & 0.523 & 0.544 \\

           & +Node2Vec & GloVe+Node2Vec & 0.553 & false & 0.558 & 0.615 & 0.585 \\
           &           &                &       & non-rumor & 0.540 & 0.407 & 0.464 \\
           &           &                &       & true & 0.646 & 0.716 & 0.679 \\
           &           &                &       & unverified & 0.449 & 0.477 & 0.463 \\
\hline

CNN\_BiLSTM & Base & GloVe & 0.513 & false & 0.365 & 0.590 & 0.451 \\
           &       &       &       & non-rumor & 0.447 & 0.469 & 0.457 \\
           &       &       &       & true & 0.872 & 0.648 & 0.744 \\
           &       &       &       & unverified & \textbf{0.656} & 0.323 & 0.432 \\

           & +Node2Vec & GloVe+Node2Vec & 0.487 & false & 0.387 & 0.590 & 0.467 \\
           &           &                &       & non-rumor & \textbf{0.705} & 0.296 & 0.417 \\
           &           &                &       & true & 0.491 & 0.770 & 0.600 \\
           &           &                &       & unverified & 0.621 & 0.277 & 0.383 \\
\hline

CNN\_BiGRU & Base & GloVe & 0.443 & false & \textbf{0.933} & 0.179 & 0.301 \\
           &       &       &       & non-rumor & 0.336 & 0.444 & 0.381 \\
           &       &       &       & true & \textbf{0.933} & 0.567 & 0.705 \\
           &       &       &       & unverified & 0.305 & 0.615 & 0.408 \\

           & +Node2Vec & GloVe+Node2Vec & 0.611 & false & 0.600 & 0.423 & 0.496 \\
           &           &                &       & non-rumor & 0.530 & 0.753 & 0.622 \\
           &           &                &       & true & 0.707 & 0.851 & 0.773 \\
           &           &                &       & unverified & 0.641 & 0.384 & 0.481 \\
\hline

\end{tabular}
\end{table*}

Other improvements were also explored to boost the model's performance.
For the CNN\_BiGRU model, we implemented a training method with cross-validation, which raised the accuracy of the model from 0.758 to 0.775. 
The cross-validation technique trains and validates the model on different subsets of data.
This provides a more reliable estimate of the model's performance.
We also experimented with a scheduler for adjusting the learning rate during training.
Although the improvements from the learning rate scheduler were almost invisible, in future work, we plan to experiment with the parameters to find a combination that could help the model.

\subsection{Fake News Mitigation}
The experimental validation shows that SparseShield has the best results in the Mitigation stage compared to NetShield and Random Solver.
This indicates its effectiveness in preventing the spread of false information within the tested parameters.
Future research could experiment with larger networks to evaluate the performance of each algorithm on larger graphs.
In addition, we plan to test other state-of-the-art algorithms, e.g., MCWDST~\cite{Truica2023MCWDST}, CONTAIN~\cite{Apostol2024contain}, to discover the best approach for mitigating false information in a real-world network.

\subsection{Limitations}
This study has several limitations that should be considered when interpreting the results:

\textbf{Limited number and size of datasets:} 
    The experiments are conducted on only two publicly available datasets, both of which are relatively small.
    This may restrict the generalizability of the findings and may limit the robustness of the conclusions.

\textbf{Unstable data sources:} 
    The datasets depend on metadata and external URLs that may change or become inaccessible over time, which presents reproducibility issues and complicates long-term research use.

\textbf{Lack of expert annotation:} 
    Not all samples in the datasets are annotated by domain experts.
    The presence of automatically labeled or weakly supervised data may introduce inconsistencies and reduce the quality of the evaluation.

\textbf{Absence of user network structure:} 
    Many existing datasets lack information about the underlying network structure of users (e.g., followers, interactions), which limits the ability to model relational or propagation-based features that are often critical in social media analysis.

\textbf{Large Language Models models not explored:} 
    While the models tested perform competitively, more recent large language models demonstrate superior performance across NLP tasks, which were not fully utilized and may offer improved results.

\textbf{Performance similar to state-of-the-art models}: 
    While the proposed method achieves similar results with state-of-the-art models, it does not outperform them significantly on the evaluated datasets, highlighting opportunities for future methodological enhancements.

In future work, we aim to address these limitations by utilizing larger and more diverse datasets with stable and well-annotated content, incorporating network information when available, and employing more advanced architectures to enhance model performance further.

\section{Conclusions}~\label{sec:conclusions}

In this paper, we propose \textsc{CleanNews}, a three-level architecture for fake news detection and mitigation, using two real-world datasets: Twitter15 and Twitter16. 
As part of our approach, we introduced a novel embedding technique that integrates both textual content and user network information, enabling the model to capture not only linguistic features but also the structural patterns associated with misinformation spread.

The \textbf{Preprocessing module} uses notable techniques to remove unnecessary information.
The module included: data cleaning, Stopword removal, tokenization, and stemming/lemmatization.
We experiment with different preprocessing techniques, comparing their influence on the accuracy of the model.
The best result was obtained with soft data cleaning, where we removed only the punctuation marks and emojis.

During the \textbf{Fake News Detection module}, we compare eight models containing BiLSTM/BiGRU and CNN layers to discover the best approach for \textsc{CleanNews}, using a grid search to extract the optimal parameters.
For the Twitter15 dataset, the CNN\_BiGRU model with DeBERTa and Node2Vec embeddings achieved the highest accuracy of $0.758$.
For the Twitter16 dataset, the  BiLSTM\_CNN model with only the DeBERTa embeddings obtained an accuracy of $0.823$.
For both datasets, the CNN\_BiGRU model with DeBERTa embeddings had the best precision when detecting false information.
We added additional improvements to the best model that uses CNN\_BiGRU with DeBERTa + Node2Vec embeddings to obtain \textsc{CleanNews}, integrating training with a cross-validation technique that increased the model's accuracy by 2\%. 
Although the results were not as good as the best results obtained on the Twitter15 and Twitter16 datasets, the models we developed outperformed similar models used on these datasets.

For the \textbf{Fake News Mitigation module}, we compared three algorithms, i.e., SparseShield, NetShield, and Random Solver, using real-world datasets to analyze their performance in preventing the spread of harmful information within the network.
Among the three, the SparseShield algorithm achieved the best results on both datasets, saving almost 50\% of the nodes from being affected during the information spread.
This significant reduction highlights the algorithm's potential for mitigating the impact of fake news on social networks.

In addition to the core model architecture and network-based mitigation strategies, our solution incorporates a comprehensive ablation study to assess the individual contributions of key model components.
This helped validate the effectiveness of each architectural choice.
Furthermore, we performed extensive hyperparameter tuning to optimize the model's performance on both datasets, ensuring a robust and well-calibrated approach to fake news detection using \textsc{CleanNews}. 

Despite the promising results, several limitations should be acknowledged.
Most publicly available fake news datasets are limited in size, often lack expert annotations, and rarely include detailed user network structures such as follower graphs or interaction histories.
This lack of comprehensive datasets poses a major challenge to capturing the full complexity of misinformation dynamics, particularly the social and relational factors that drive how false information spreads across networks.
While the study explored several deep learning architectures, it did not fully leverage newer transformer-based models beyond DeBERTa, which may offer further performance gains.
Finally, although our approach demonstrated competitive performance, it did not significantly outperform existing state-of-the-art methods, highlighting room for future enhancements.

For future research directions, we aim to analyze the best strategy for preprocessing and using other state-of-the-art embeddings like BERT, RoBERTa, and GPT-based models.
We also plan to experiment with a variety of parameters on each model, adding other techniques like cross-validation on each of them, as well as experiment with other Network Immunization algorithms on larger graphs.

\section*{Acknowledgment}

The research presented in this article was supported in part by
The Academy of Romanian Scientists, through the funding of the projects 
1) ``SCAN-NEWS: Smart system for deteCting And mitigatiNg misinformation and fake news in social media'' (AOȘR-TEAMS-III) and 
2) ``NetGuardAI: Intelligent system for harmful content detection and immunization on social networks'' (AOȘR-TEAMS-IV).

\bibliographystyle{elsarticle-num-names} 
\bibliography{main}

\end{document}